\documentclass[aps,prd,nofootinbib,floatfix,superscriptaddress,secnumarabic]{revtex4}

\usepackage{xcolor}
\usepackage{hyperref}
\usepackage{amsmath}
\usepackage{bm}
\usepackage{graphicx}
\usepackage{epsf}
\usepackage{epsfig}
\usepackage{amsmath}
\usepackage{amsfonts,amssymb}
\usepackage{colordvi}
\usepackage{color}
\usepackage{lscape}
\usepackage{rotfloat}
\usepackage{rotating}
\usepackage{dsfont}
\usepackage{slashed,ulem}
\usepackage{booktabs}
\usepackage{rotating}
\usepackage{comment}
\usepackage{hhline}
\usepackage{subcaption}
\newcommand{\be}{\begin{equation}}
\newcommand{\ee}{\end{equation}}
\newcommand{\bea}{\begin{eqnarray}}
\newcommand{\eea}{\end{eqnarray}}

\newcommand{\MSbar}{{\overline{\rm MS}}}
\newcommand{\gtilde}{\frac{g^2}{16 \, \pi^2}\; }

\newcommand{\pa}{\partial}
\newcommand{\la}{\lambda}

\def\slashed{{/}\mskip-10.0mu}

\begin{document}

\title{One-loop mixing matrix of the ${\cal N}{=}1$ SYM Supercurrent operator upon renormalization on the Lattice using stout-smeared Links}

\author{M.~Costa}
\email[]{marios.costa@cut.ac.cy}
\affiliation{Department of Mechanical Engineering and Material Science and Engineering, Cyprus University of Technology, Limassol, CY-3036, Cyprus}

\author{E.~Antoniou}
\email[]{elias.a@rinnoco.com}
\affiliation{Rinnoco Ltd, Limassol, CY-3047, Cyprus}

\author{C.~Costa}
\email[]{costa.c@rinnoco.com}
\affiliation{Rinnoco Ltd, Limassol, CY-3047, Cyprus}

\author{S.~Dimou}
\email[]{salomi@rinnoco.com}
\affiliation{Rinnoco Ltd, Limassol, CY-3047, Cyprus}

\author{H.~Panagopoulos}
\email[]{panagopoulos.haris@ucy.ac.cy}
\affiliation{Department of Physics, University of Cyprus, Nicosia, CY-1678, Cyprus}

\begin{abstract}
We extend the perturbative investigation of the renormalization and mixing patterns of the supercurrent operator in lattice ${\cal N}=1$ supersymmetric Yang-Mills (SYM) theory, using stout-smeared gauge links. We employ the SLiNC action for gluinos and a family of Symanzik improved gauge actions for gluons. Building upon previous studies employing clover-improved fermions and the Wilson gauge action, we compute the one-loop mixing matrix of the supercurrent operator among gauge-invariant and gauge-variant operators. Particular emphasis is placed on the structure of operator mixing in the presence of stout smearing parameters. Our results contribute toward refining fine-tuning strategies necessary for the restoration of supersymmetry in the continuum limit, and provide a stepping stone for future perturbative and nonperturbative investigations in supersymmetric theories on the lattice.
\end{abstract}

\maketitle
\section{Introduction}

Supersymmetry (SUSY) plays a central role in modern particle physics. Beyond its elegant unification of bosonic and fermionic degrees of freedom, SUSY not only offers compelling solutions~\cite{ParticleDataGroup:2026mpi} to outstanding problems such as the hierarchy problem, but also provides several promising candidates for dark matter~\cite{Jungman:1995df, ParticleDataGroup:2026mpi} within extensions of the Standard Model. Furthermore, Quantum Field Theories (QFTs) with extended supersymmetry, such as ${\cal N}=4$ supersymmetric Yang-Mills (SYM) theory, serve as important testing grounds for the gauge/gravity duality and for understanding the dynamics of strongly coupled gauge theories.

A particularly appealing setting for studying nonperturbative aspects of SUSY is the ${\cal N}=1$ SYM theory, which describes the interactions of gluons and their fermionic superpartners, the gluinos. This theory retains many of the essential features of supersymmetric gauge theories involving matter fields, while being simpler and more amenable to precise numerical simulations. Despite these advantages, the formulation of supersymmetric theories on the lattice faces significant challenges. Lattice discretization necessarily breaks supersymmetry at finite lattice spacing~\cite{Ali:2019agk, Bergner:2015adz, Feo:2002yi, Kaplan:2007zz, Catterall:2009it}, due to the loss of continuous translation symmetry and the impossibility of preserving the Leibniz rule for lattice derivatives. Recovering supersymmetry in the continuum limit requires careful fine-tuning, guided by SUSY Ward identities~\cite{Farchioni:2001wx, Feo:2003km}. However, the use of Ward identities~\cite{Taniguchi:1999fc, Ali:2018fbq} for this purpose is complicated by operator mixing, since the divergence of the supercurrent mixes with a variety of gauge-invariant and gauge-variant operators upon renormalization.

In a previous study~\cite{Bergner:2022wnb}, we performed a detailed one-loop perturbative calculation of the renormalization and mixing of the supercurrent operator in ${\cal N}=1$ SYM. Using clover-improved Wilson fermions and the standard Wilson gauge action, we computed two- and three-point Green's functions of the supercurrent with external gluon and gluino fields, evaluated in an arbitrary covariant gauge. We systematically constructed the full basis of operators up to dimension~$7/2$, respecting global symmetries such as Lorentz invariance, ghost number conservation, and charge conjugation symmetry. The resulting mixing pattern was encoded in a $13 \times 13$ block-triangular matrix, and the renormalization factors were determined both in dimensional and lattice regularizations, employing the $\MSbar$ scheme.

This previous work established a systematic perturbative framework for analyzing the operator mixing of the supercurrent on the lattice and provided a basis for future nonperturbative determinations of the mixing coefficients via SUSY Ward identities. It also highlighted the importance of gauge-invariant renormalization schemes, such as the Gauge-Invariant Renormalization Scheme (GIRS)~\cite{Costa:2021pfu, Costa:2021iyv}, where only gauge-invariant operators contribute, dramatically simplifying the mixing structure.

The aim of the present work is to extend and refine previous investigations by performing one-loop calculations of renormalization and mixing coefficients, using improved versions of the lattice action and of the supercurrent. In particular, this includes the incorporation of stout-smeared gauge links both in the covariant derivatives of the gluino action and in the lattice definitions of the operators. In addition, we employ three Symanzik-improved gluon actions that are commonly used in simulations. Our choice to apply stout smearing only to the covariant derivatives follows the SLiNC (Stout Link Non-perturbative Clover) strategy adopted in lattice QCD~\cite{Cundy:2009yy, 
Horsley:2008fw, 
Horsley:2008ap}: The clover term remains constructed from unsmeared links, as these already span paths of length $4 a$ in all possible orientations ("$a$" stands for the lattice spacing). A word of caution here is that gauge fluctuations originating from the clover term have a large impact on numerical stability of QCD simulations~\cite{Francis:2019muy}; nevertheless, an attempt to reduce these fluctuations by smearing the links in the clover term has the downside of introducing terms in the action with a very pronounced spatial extent, leading to an undesirably extended (less sparse) fermion matrix.

The structure of the paper is as follows. In Section~\ref{previous_summary}, we summarize the theoretical and computational framework for the renormalization of the supercurrent, with further details available in Ref.~\cite{Bergner:2022wnb}. Section~\ref{LatticeSetup} introduces the lattice formulation adopted in this work, while Section~\ref{Latt} presents our new lattice results for the one-loop mixing matrix of the supercurrent operator upon renormalization, computed using lattice regularization and expressed in the $\MSbar$ scheme. Finally, Section~\ref{summary} offers a brief summary and an outlook on future directions, with particular emphasis on extensions to supersymmetric QCD. Appendix~\ref{app:MSbarGreenFunctions} contains all the necessary $\MSbar$-renormalized Green’s functions required to compute the mixing matrix of the supercurrent operator in the $\MSbar$ scheme. 

\section{Computational Setup}
\label{previous_summary}

In our previous work~\cite{Bergner:2022wnb}, we established the computational framework for the perturbative renormalization of the supercurrent operator in lattice ${\cal N}=1$ supersymmetric Yang-Mills (SYM) theory~\cite{Veneziano:1982ah, Curci:1986sm}. For a detailed exposition of the methodology and underlying formalism, we refer the reader to Ref.~\cite{Bergner:2022wnb}. Following Noether’s procedure, the expression for the supercurrent operator~\cite{Martin:1997ns, Gates:1983nr} takes the form:
\begin{equation}
S_\mu = -\frac{1}{2} \, \mathrm{tr}_c \left( u_{\rho\sigma} \, [\gamma_\rho, \gamma_\sigma] \, \gamma_\mu \lambda \right),
\end{equation}
where $u_{\mu \nu}$ denotes the gluon field tensor; $u_{\mu \nu} = \pa_{\mu}u_{\nu} - \pa_{\nu}u_{\mu} + i g\, [u_{\mu},u_{\nu}]$, $u_{\mu}$ is the gluon field and $\lambda$ is the gluino field. A proper study of $S_\mu$ must account for the fact that, under renormalization, it mixes with a variety of other operators carrying the same quantum numbers~\cite{ Curci:1986sm, Bergner:2022wnb}. Among these operators, there is only one  gauge-invariant operator:
\begin{equation}
T_\mu = 2\, \mathrm{tr}_c \left( u_{\mu\nu} \, \gamma_\nu \, \lambda \right).
\end{equation}
Furthermore, $S_\mu$ being a composite operator, it may mix with other operators  which share the same global symmetries (e.g., Lorentz symmetry or hypercubic symmetry on the lattice, global $SU(N_c)$ transformations, ghost number conservation) and whose dimension is less than or equal to that of the supercurrent operator. These mixing operators are constructed from gluino, gluon, and ghost fields. 
From general renormalization theorems these operators fall into four classes~\cite{Collins:1984xc}:
\begin{itemize}
  \item \textbf{Class G: Gauge-invariant operators.} \\
  This class only includes $T_\mu$, which mixes with $S_\mu$, and vice versa. No other gauge-invariant operators contribute to the mixing within this class.

  \item \textbf{Class A: BRST-exact operators.} \\
  These arise as variations under the Becchi-Rouet-Stora-Tyutin (BRST) symmetry~\cite{Suzuki:2012pc}. An explicit case is:
  \begin{equation}
    \mathcal{O}_{A1} = \frac{1}{\alpha} \, \mathrm{tr}_c \left( (\partial_\nu u_\nu)\, \gamma_\mu \lambda \right)
    - i g\, \mathrm{tr}_c \left( [c, \bar{c}]\, \gamma_\mu \lambda \right),
  \end{equation}
  where $\alpha$ is the gauge-fixing parameter, and $c$, $\bar{c}$ denote the ghost and anti-ghost fields, respectively.

  \item \textbf{Class B: Operators vanishing by the equations of motion.} \\
  These vanish upon using the classical equations of motion:
  \begin{align}
    \mathcal{O}_{B1} &= \mathrm{tr}_c \left( u_\mu\, \slashed{D} \lambda \right), \\
    \mathcal{O}_{B2} &= \mathrm{tr}_c \left( \slashed{u}\, \gamma_\mu\, \slashed{D} \lambda \right),
  \end{align}
where 
${\cal{D}}_\mu \lambda =  \pa_{\mu} \lambda + i g \,[u_{\mu},\lambda]$. 
  \item \textbf{Class C: Operators compatible with global symmetries but not belonging to classes G, A, B.} \\
  These operators are allowed by the global symmetries of the theory and can mix with $S_\mu$ through finite renormalization coefficients. They must be included in the complete mixing analysis
  :
  \begin{align}
    \mathcal{O}_{C1} &= \mathrm{tr}_c \left( u_\mu\, \lambda \right), \\
    \mathcal{O}_{C2} &= \mathrm{tr}_c \left( \slashed{u}\, \gamma_\mu\, \lambda \right), \\
    \mathcal{O}_{C3} &= \mathrm{tr}_c \left( \slashed{u}\, \partial_\mu \lambda \right), \\
    \mathcal{O}_{C4} &= \mathrm{tr}_c \left( (\partial_\mu \slashed{u})\, \lambda \right), \\
    \mathcal{O}_{C5} &= \mathrm{tr}_c \left( (\partial_\nu u_\nu)\, \gamma_\mu \lambda \right), \\
    \mathcal{O}_{C6} &= \mathrm{tr}_c \left( u_\nu\, \gamma_\mu\, \partial_\nu \lambda \right), \\
    \mathcal{O}_{C7} &= i g\, \mathrm{tr}_c \left( [u_\rho, u_\sigma]\, [\gamma_\rho, \gamma_\sigma]\, \gamma_\mu \lambda \right), \\
    \mathcal{O}_{C8} &= i g\, \mathrm{tr}_c \left( [u_\mu, u_\nu]\, \gamma_\nu \lambda \right). 
  \end{align}
\end{itemize}


[In principle, some Lorentz-noncovariant (but hypercubic-covariant) operators can appear in Class C, namely operators analogous to $\mathcal{O}_{C3} -\mathcal{O}_{C7}$, with 3 Lorentz indices being set equal to $\mu$, e.g.:
$\mathcal{O}_{C3} = \mathrm{tr}_c \left( \slashed{u}\, \partial_\mu \lambda \right) \to \mathrm{tr}_c \left(u_\mu \gamma_\mu\, \partial_\mu \lambda \right)$. However, it turns out that no such operators appear in the one-loop Green's functions, so, at least to one loop, no such mixing is present.]

Therefore, the renormalized supercurrent $S_\mu^R$ is expressed as a linear combination:
\begin{equation}
S_\mu^R = Z_{SS} S_\mu^B + Z_{ST} T_\mu^B + Z_{SA_1} {\cal O}_{A_1}^B
+ \sum_{i=1}^{2} Z_{SB_i} {\cal O}_{B_i}^B
+ \sum_{i=1}^{8} Z_{SC_i} {\cal O}_{C_i}^B,
\label{Zz1}
\end{equation}
where the superscript $B$ ($R$) denotes bare (renormalized) operators, and $Z_{{\cal O}{\cal O'}}$ indicate the coefficient with which  operator ${\cal O'}$ mixes into the renormalized version of operator ${\cal O}$. The coefficients $Z_{\cal O O'}$ should also be properly labeled by the
regularization, $X$, and the renormalization scheme, $Y$, which are being used,
and are therefore written as $Z^{XY}_{\cal O O'}$.
Eq.~\eqref{Zz1} 
is the first row of the mixing matrix, whose matrix elements we calculate in the lattice
regularization scheme ($LR$) at one-loop order and in the $\MSbar$ scheme.

Perturbative one-loop computations of two-point and three-point Green's functions with external gluon, gluino, and ghost fields are performed in an arbitrary covariant gauge. The relevant field renormalization constants are:
\begin{align}
u_\mu^R &= \sqrt{Z_u}\, u_\mu^B, \\
\lambda^R &= \sqrt{Z_\lambda}\, \lambda^B, \\
c^R &= \sqrt{Z_c}\, c^B, \\
g^R &= Z_g \mu^{-\epsilon} g^B,
\end{align}
where $\mu$ is the dimensional regularization scale and $\epsilon$ is related to the spacetime dimension by $d=4-2\epsilon$; the quantity $\mu$ is connected to the standard $\MSbar$ renormalization scale, $\bar \mu$, via: $\mu = \bar\mu \, (4\pi/e^{\gamma_E})^{-1/2}$ ($\gamma_E$ is the Euler constant).

One important outcome of this study is the ratio of the renormalization constants of the supercurrent $S_\mu$ and its gauge-invariant mixing partner $T_\mu$, namely $Z_{SS}/Z_{ST}$. These operators are of central importance because of their role in realizing supersymmetry at the quantum level. To obtain a comprehensive understanding of the mixing structure, we also determine the lattice mixing coefficients of $S_\mu$ with BRST-invariant operators, equation-of-motion operators, and finite-mixing operators. The coefficients $Z^{LR,\MSbar}_{\cal O O'}$ are all determined in this work using the action in Eq.~\eqref{susylagrLattice}.

The continuum computations, identical to those presented in our earlier work~\cite{Bergner:2022wnb}, are carried out using dimensional regularization and yield the corresponding renormalized Green’s functions in the $\overline{\text{MS}}$ scheme. To keep this work self-contained, these results are provided in Appendix~\ref{app:MSbarGreenFunctions}. The continuum expressions are subsequently matched to their counterparts computed in lattice regularization. The comparison between the corresponding Green's functions allows us to extract the one-loop renormalization constants and mixing coefficients for $S_\mu$ on the lattice and in the $\MSbar$ scheme. Both the continuum and lattice calculations are technically demanding and conceptually subtle. They require a careful selection of appropriate Green’s functions and external momenta $(q_1,\ q_2,\ q_3$, see Green's functions~(\ref{2ptGF}, \ref{3ptGF}) below) that enable an unambiguous determination of the renormalization and mixing coefficients. To this end, we compute a set of two-point and three-point Green’s functions involving $S_\mu$. The two-point function is:
\begin{align}
\langle u_\nu(q_1) \, S_\mu \, \bar{\lambda}(q_2) \rangle, 
\label{2ptGF}\end{align}
and the three-point functions to be evaluated are:
\begin{align}
\langle u_\nu(q_1) \, u_\rho(q_2) \, S_\mu \, \bar{\lambda}(q_3) \rangle, \quad
\langle c(q_2) \, \bar{c}(q_1) \, S_\mu \, \bar{\lambda}(q_3) \rangle. 
\label{3ptGF}\end{align}

\begin{figure}[ht!]
\centering
\includegraphics[scale=0.3]{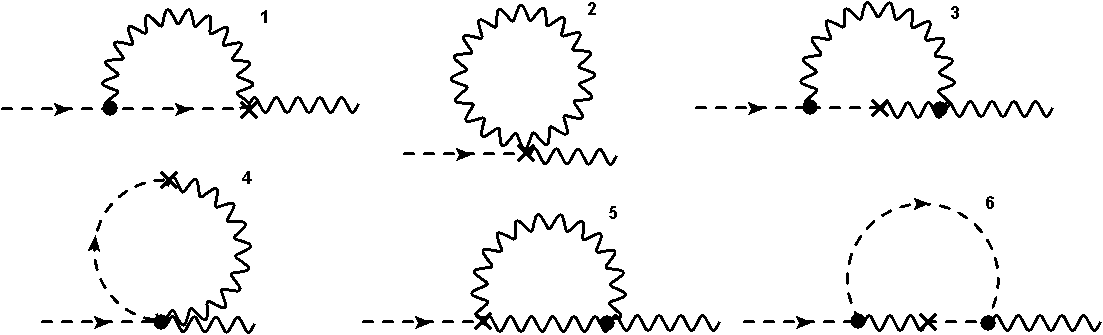}
\caption{One-loop Feynman diagrams contributing to the two-point Green's function $\langle u_\nu S_\mu \bar \la  \rangle$.  A wavy (dashed) line represents gluons (gluinos). A cross denotes the insertion of $S_\mu$. Diagrams 2, 4 do not appear in dimensional regularization; they do however show up in the lattice formulation.}
\label{fig2pt}
\end{figure}

\begin{figure}[ht!]
\centering
\includegraphics[scale=0.3]{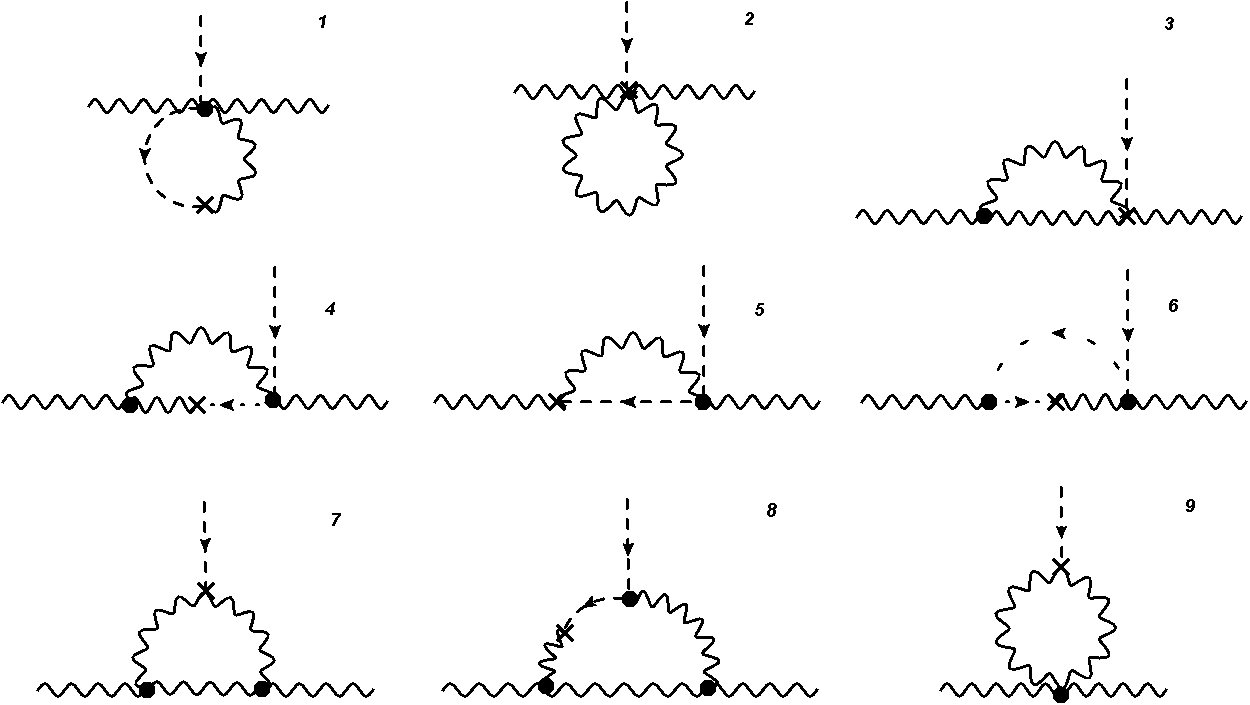}\\
\includegraphics[scale=0.3]{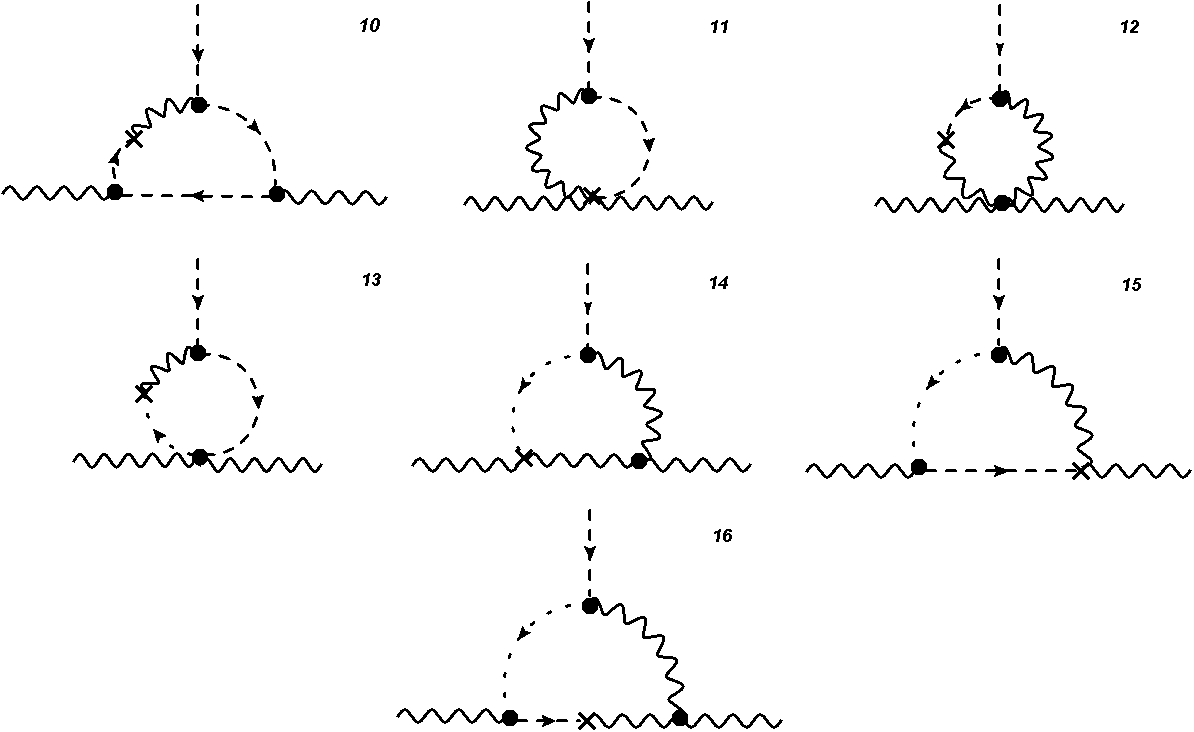}
\caption{One-loop Feynman diagrams contributing to the three-point Green's function $\langle u_\nu u_\rho S_\mu\bar \la  \rangle$.  A wavy (dashed) line represents gluons (gluinos). Diagrams 1, 2, 3, 4, 5, 6, 11, and 13 do not appear in dimensional regularization but they contribute in the lattice regularization. A cross denotes the insertion of the operator. A mirror version (under exchange of the two external gluons) of diagrams 3, 4, 5, 6, 8, 10, 14, 15 and 16 must also be included.}
\label{fig3ptguu}
\end{figure}

\begin{figure}[ht!]
\centering
\includegraphics[scale=0.3]{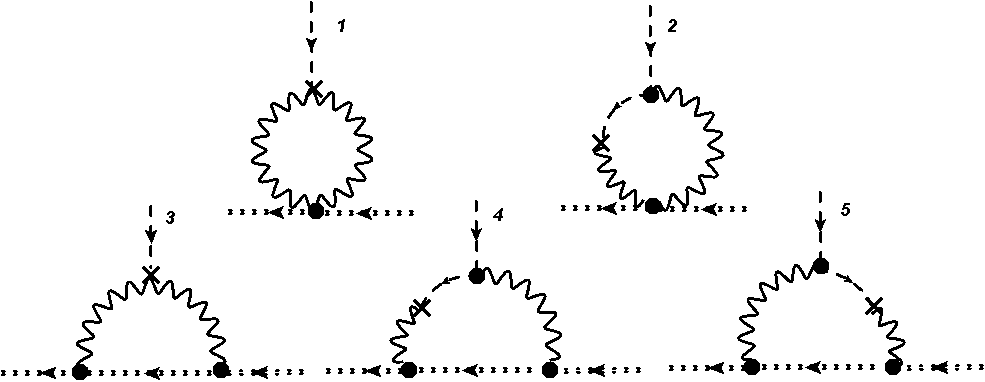}
\caption{One-loop Feynman diagrams contributing to the three-point Green's function $\langle c \, S_\mu \, \bar c \, \bar \la \rangle$.  A wavy (dashed) line represents gluons (gluinos). A cross denotes the insertion of the operator. The ``double dashed'' line is the ghost field. Diagrams 1 and 2 do not appear in dimensional regularization; they do however show up in the lattice formulation.}
\label{fig3ptgCC}
\end{figure}

Each Green’s function is expressed as a sum of Feynman diagrams; see Figs.~\ref{fig2pt}, \ref{fig3ptguu}, and \ref{fig3ptgCC}. In order to simplify the calculation, we set to zero one of the external momenta, corresponding to an external gluon, gluino or ghost field. This prescription avoids the appearance of spurious infrared singularities, and allows us to identify the operator structures that contribute to the one-loop bare Green’s functions, thereby determining the corresponding mixing coefficients. 

Even at one-loop order, several non-trivial technical challenges arise. Notably, the generation of lattice vertices involving stout-smeared gauge links and up to four gluon fields is quite intricate~\cite{Ammer:2024hqr}. However, a major advantage of these new calculations, which use stout-smeared gauge links, is that they significantly reduce discretization artifacts. This computational setup lays the groundwork for systematic studies of supersymmetry restoration in lattice ${\cal N}=1$ SYM, and opens the path toward extending the analysis to more complex theories such as supersymmetric QCD.

The tree-level Green's functions of the mixing operators can naturally appear as contributions to the corresponding one-loop bare Green's functions of the supercurrent $S_\mu$. 
To facilitate the identification of these contributions, Table~\ref{tb:treelevels} lists the tree-level expressions for the two-point and three-point Green's functions with external gluino, gluon, and ghost fields, involving the operators $S_\mu$, $T_\mu$, and all gauge-noninvariant operators that may mix with them. 

Given the tree-level expressions in Table~\ref{tb:treelevels}, our strategy is to first compute the one-loop three-point Green's function of $S_\mu$ with external gluino, ghost, and antighost fields. This calculation allows us to determine the mixing coefficient of $\mathcal{O}_{A1}$. We then proceed to the one-loop two-point Green's function of $S_\mu$ with external gluon and gluino fields. A calculation for generic values of the external momenta, $(q_1)$ and $(q_2)$, leads to very lengthy expressions involving polylogarithms of the momenta. However, to extract all renormalization and mixing coefficients appearing in Eq.~(\ref{Zz1}), it is sufficient
to consider the following three choices of values for the external momenta: ($q_2 = 0$), or ($q_1 = 0$), or ($q_2 = - q_1$). These two-point Green's functions are sufficient to determine the mixing coefficients of all other operators whose tree-level two-point Green's functions are nonvanishing. Finally, we compute the one-loop three-point Green's function of $S_\mu$ with two external gluons and one external gluino. This calculation enables us to determine the renormalization coefficients of the remaining operators, namely  $\mathcal{O}_{C7}$ and $\mathcal{O}_{C8}$.

\begin{table}[h!]
\centering
\begin{tabular}{|c | c | c | c|}
\hline
 & Tree-Level Two-Point & Tree-Level Three-Point & Tree-Level Three-Point \\
Operator $\mathcal{O}_i$ & $\ \langle u_\nu^{\alpha_1}(-q_1)\,{\cal O}_{i}(x)\, \bar\lambda^{\alpha_2}(q_2) \rangle\ $ & $\ \langle u_\nu^{\alpha_1}(-q_1)\, u_\rho^{\alpha_2}(-q_2)\,{\cal O}_{i}(x)\, \bar\lambda^{\alpha_3}(q_3) \rangle\ $ & $\ \langle c^{\alpha_3}(q_3) \,{\cal O}_{i}(x)\,\bar c^{\alpha_2}(q_2) \bar\lambda^{\alpha_1}(q_1) \rangle\ $ \\[0.5ex]
\hhline{|=|=|=|=|}
$S_\mu$ & $-i (\slashed{q_1} \gamma_\nu - q_{1\nu}) \gamma_\mu$ & $g [\gamma_\nu,\gamma_\rho] \gamma_\mu / 2$ & $0$ \\[2ex]
\hline
$T_\mu$ & $i (q_{1\mu} \gamma_\nu - \slashed{q_1} \delta_{\mu \nu})$ & $-g (\delta_{\mu \nu} \gamma_\rho - \delta_{\mu \rho} \gamma_\nu)$ & $0$ \\[2ex]
\hline
${\cal O}_{A1}$ & $i q_{1\nu} \gamma_\mu / (2\alpha)$ & $0$ & $(g/2) \gamma_\mu$ \\[2ex]
\hline
${\cal O}_{B1}$ & $i \delta_{\mu \nu} \slashed{q_2} / 2$ & $-g (\delta_{\nu \mu} \gamma_\rho - \delta_{\rho \mu} \gamma_\nu) / 2$ & $0$ \\[2ex]
\hline
${\cal O}_{B2}$ & $i \gamma_\nu \gamma_\mu \slashed{q_2} / 2$ & $\, g (\gamma_\nu \gamma_\rho \gamma_\mu - \delta_{\rho\nu}\gamma_\mu + \delta_{\mu\nu}\gamma_\rho  - \delta_{\mu\rho} \gamma_\nu) $ & $0$ \\[2ex]
\hline
${\cal O}_{C1}$ & $\delta_{\mu \nu} / 2$ & $0$ & $0$ \\[2ex]
\hline
${\cal O}_{C2}$ & $\gamma_\nu \gamma_\mu / 2$ & $0$ & $0$ \\[2ex]
\hline
${\cal O}_{C3}$ & $i \gamma_\nu q_{2\mu} / 2$ & $0$ & $0$ \\[2ex]
\hline
${\cal O}_{C4}$ & $i \gamma_\nu q_{1\mu} / 2$ & $0$ & $0$ \\[2ex]
\hline
${\cal O}_{C5}$ & $i \gamma_\mu q_{1\nu} / 2$ & $0$ & $0$ \\[2ex]
\hline
${\cal O}_{C6}$ & $i \gamma_\mu q_{2\nu} / 2$ & $0$ & $0$ \\[2ex]
\hline
${\cal O}_{C7}$ & $0$ & $-g [\gamma_\nu, \gamma_\rho] \gamma_\mu$ & $0$ \\[2ex]
\hline
${\cal O}_{C8}$ & $0$ & $-g (\delta_{\nu \mu} \gamma_\rho - \delta_{\rho \mu} \gamma_\nu) / 2$ & $0$ \\[2ex]
\hline
\end{tabular}
\caption{The two-point and three-point tree-level Green's functions of $S_\mu$ and $T_\mu$ as well as of gauge noninvariant operators which may mix with $S_\mu$\,.
$\langle u_\nu^{\alpha_1}(-q_1)\,{\cal O}_{i}(x)\, \bar\lambda^{\alpha_2}(q_2) \rangle_{amp}^{tree}$ and $\langle u_\nu^{\alpha_1}(-q_1)\, u_\rho^{\alpha_2}(-q_2)\,{\cal O}_{i}(x)\, \bar\lambda^{\alpha_3}(q_3) \rangle_{amp}^{tree}$ are shown apart from an overall factor of $\delta^{\alpha_{1}\alpha_{2}} e^{i\,x\cdot(q_1+q_2)}$ and $f^{\alpha_{1}\alpha_{2}\alpha_{3}} e^{i\,x\cdot(q_1+q_2+q_3)}$, respectively. Similarly, $\langle c^{\alpha_3}(q_3) \,{\cal O}_{i}(x)\,\bar c^{\alpha_2}(q_2) \bar\lambda^{\alpha_1}(q_1) \rangle_{amp}^{tree}$ is shown apart from an overall factor of $f^{\alpha_{1}\alpha_{2}\alpha_{3}} e^{i\,x\cdot(q_1-q_2+q_3)}$.
\label{tb:treelevels}}
\end{table}

In the following section, we proceed to outline the lattice formulation and then we present the calculation of all relevant Green's functions using lattice regularization. Our results, taken together with the results in $DR$, which are included Appendix~\ref{app:MSbarGreenFunctions}, allow us to deduce all renormalization and mixing coefficients in $\MSbar$.

\newpage 
\section{Lattice Perturbation Theory and Results}
\label{resultsALL}

\subsection{Formulation on the lattice}
\label{LatticeSetup}
In this work, we perform a complete one-loop determination of the mixing pattern of the supercurrent operator through perturbative calculations, using gauge-variant, off-shell Green’s functions and a lattice discretization based on stout smearing. To ensure general applicability, we explicitly distinguish between the stout-smearing parameters used in the fermion action and those appearing in the definition of the composite operator $S_\mu$. This distinction enables a more flexible computational framework and supports a systematic investigation of discretization effects and operator improvement. In Refs.~\cite{Bergner:2022wnb, Costa:2020keq}, we investigated the mixing of various composite operators under renormalization in SYM, including the Supercurrent operator and the Gluino-Glue, using Wilson and clover fermions. In these works, as well as in our previous lattice calculations for supersymmetric QCD (SQCD)~\cite{Costa:2017rht, Costa:2018mvb, Costa:2023cqv, Costa:2024tyz}, standard discretization was employed;  gluinos are defined on the lattice sites, while gluons are represented by link variables:  
\begin{equation}
U_\mu (x) \equiv U_{x,x+\mu} = e^{i g a T^{\alpha} u_\mu^\alpha (x+a\hat{\mu}/2)},
\end{equation}
where $\alpha$ is the color index in the adjoint representation of the gauge group $SU(N_c)$. This formulation explicitly breaks all SUSY generators and chiral symmetry, necessitating fine-tuning in numerical simulations of SQCD. 

To improve chiral symmetry restoration at finite lattice spacing and to reduce the number of required fine-tunings, we now employ stout-smeared links, $\Tilde{U}_\mu$. The stout smearing procedure enhances the behavior of operator matrix elements by systematically reducing ultraviolet fluctuations in the gauge fields~\cite{Morningstar:2003gk}. It is typically applied iteratively, starting from the original link variables $U_\mu$, which are modified according to the transformation~\cite{Morningstar:2003gk}, \cite{Horsley:2008ap}:
\begin{equation}
\Tilde{U}_\mu(x) = e^{i\,Q_\mu(x)}\, U_\mu(x)\,,
\end{equation}  
where $Q_\mu(x)$ is given by:
\begin{equation}
  Q_\mu(x) = \frac{\omega}{2\,i} \left[V_\mu(x) U_\mu^\dagger(x) -  
  U_\mu(x)V_\mu^\dagger(x) -\frac{1}{3} \text{Tr} \,\Big(V_\mu(x)  
  U_\mu^\dagger(x) -  U_\mu(x)V_\mu^\dagger(x)\Big)\right] \, .
\label{Q_def}
\end{equation}  
Here, the ``stout coefficient'' $\omega$ is a tunable parameter, while $V_\mu(x)$ represents the sum over all staples associated with the link $U_\mu(x)$.

In our lattice formulation, gluinos are represented by clover-improved Wilson fermions in the adjoint representation; stout smearing is applied to the link variables present in the covariant derivatives, but not to the clover term\footnote{Our use of stout smearing is restricted to the links entering the covariant derivatives, following the SLiNC strategy employed in lattice QCD~\cite{Cundy:2009yy,Horsley:2008fw,Horsley:2008ap}. The clover term is constructed from unsmeared links, which already sample paths of length $4a$ in all possible orientations. Although gauge fluctuations associated with the clover term can significantly affect the numerical stability of QCD simulations~\cite{Francis:2019muy}, smearing its links would introduce terms with a much larger spatial extent, making the fermion matrix less sparse. 
}. 
For the gauge fields we employ a class of Symanzik improved gauge actions~\cite{Horsley:2008ap}, involving Wilson loops with 4 and 6 links ($1\times1$ plaquettes and $1\times2$ rectangles, respectively). The Euclidean action ${\cal S}^{LR}_{\rm SYM}$ on the lattice then takes the form:
\begin{equation}
\begin{split}
{\cal S}^{LR}_{\rm SYM} \;=\;  \sum_{x} \Bigg\{ 
& \frac{2}{g^2} \Bigg[ 
   c_0 \sum_{\rm plaq} \, {\rm Re\,Tr}\!\left( 1 - U_{\rm plaq} \right) 
 + c_1 \sum_{\rm rect} \, {\rm Re\,Tr}\!\left( 1 - U_{\rm rect} \right) 
\Bigg] \\[6pt]
& + a^{4} \sum_{\mu} \Bigg[
   {\rm Tr}\!\left( \bar\lambda \gamma_{\mu} \Tilde{D}_{\mu} \lambda \right)
 - \frac{a r}{2} \, {\rm Tr}\!\left( \bar\lambda \Tilde{D}^{2} \lambda \right)
\Bigg] - a^{4}\sum_{\mu,\nu} \Bigg[
   \frac{c_{\rm SW}\,a}{4} \, 
   \bar\lambda^{\alpha} \sigma_{\mu \nu} \hat{F}_{\mu \nu}^{\alpha \beta} \lambda^{\beta}
\Bigg]
+  m_0 \, {\rm Tr}\!\left( \bar\lambda \lambda \right) \Bigg \},
\end{split}
\label{susylagrLattice}
\end{equation}
where  the plaquette ($U_{\rm plaq}$) and rectangle ($U_{\rm rect}$) loop variables are given by:
\begin{align}
U_{\rm plaq}(x;\mu,\nu) &= U_\mu(x)\, U_\nu(x+a\hat\mu)\, 
U_\mu^\dagger(x+a\hat\nu)\, U_\nu^\dagger(x), \\
U_{\rm rect}(x;\mu,\nu) &= U_\mu(x)\, U_\mu(x+a\hat\mu)\, 
U_\nu(x+2a\hat\mu)\, U_\mu^\dagger(x+a\hat\nu+a\hat\mu)\,
U_\mu^\dagger(x+a\hat\nu)\, U_\nu^\dagger(x),
\end{align}
the coefficients $c_0$ and $c_1$ can in principle be chosen arbitrarily, subject to the following normalization condition, which ensures the correct classical continuum limit of the action:
\begin{equation}
c_0 + 8 c_1  = 1
\label{norm}
\end{equation}
For the numerical evaluation, particular choices of values for $\{c_0$, $c_1\}$ are employed in our calculations as shown in Table \ref{tb:Symanzik_coeff}.
\begin{table}[h!]
\centering
\renewcommand{\arraystretch}{1.15}
\begin{tabular}{|c|c|c|c|}
\hline
Gluon action & Wilson & Tree-level Symanzik (TLS) & Iwasaki \\
\hhline{|=|=|=|=|}
$c_{0}$ & $1$ & $5/3$ & $3.648$ \\ \hline
$c_{1}$ & $0$ & $-1/12$ & $-0.331$ \\
\hline
\end{tabular}
\caption{Commonly used sets of values for the Symanzik coefficients.}
\label{tb:Symanzik_coeff}
\end{table}

The 4-vector $x$ is restricted to the values $x = na$, with $n$ being an integer 4-vector. The terms proportional to the Wilson parameter, $r$, eliminate the problem of fermion doubling, at the expense of breaking chiral invariance\footnote{In what follows, we will set $|r|=1$.}. In the limit $a \to 0$ the classical lattice action reproduces the continuum one. A gauge-fixing term, together with the compensating ghost field term, must also be added to the action, in order to avoid divergences from the integration over gauge orbits; these terms are the same as in the non-supersymmetric case~\cite{Kawai:1980ja}. Similarly, a standard ``measure'' term must also be included, in order to account for the Jacobian in the change of integration variables: $U_\mu \to u_\mu$\,. 

The definitions of the covariant derivatives are as follows:
\bea
{\Tilde{\cal{D}}}_\mu\lambda(x) &\equiv& \frac{1}{2a} \Big[ \Tilde{U}_\mu (x) \lambda (x +  a \hat{\mu}) \Tilde{U}_\mu^\dagger (x) - \Tilde{U}_\mu^\dagger (x - a \hat{\mu}) \lambda(x - a \hat{\mu}) \Tilde{U}_\mu(x - a \hat{\mu}) \Big], \\
{\cal D}^2 \lambda(x) &\equiv& \frac{1}{a^2} \sum_\mu \Big[ \Tilde{U}_\mu (x)  \lambda (x + a \hat{\mu}) \Tilde{U}_\mu^\dagger (x)  - 2 \lambda(x) +  \Tilde{U}_\mu^\dagger (x - a \hat{\mu}) \lambda (x - a \hat{\mu}) \Tilde{U}_\mu(x - a \hat{\mu})\Big].
\eea

The definition of $\hat{F}_{\mu \nu}^{\alpha \beta}$ in the adjoint representation is:

\bea
\hat{{F}}_{\mu \nu}^{\alpha \beta} \! \! &=& \! \! \frac{1}{8}({Q}_{\mu \nu}^{\alpha \beta}-{Q}_{\nu \mu}^{\alpha \beta}), \qquad \sigma_{\mu\nu}=\frac{1}{2}[\gamma_\mu,\gamma_\nu],\\ {\textrm{where:\quad}}
{Q}_{\mu \nu}^{\alpha \beta} \! \! &=& \! \! 2{\rm{tr}}_c \bigg( 
T^\alpha\, U_{x,x+\mu}U_{x+\mu,x+\mu+\nu}U_{x+\mu+\nu,x+\nu}U_{x+\nu,x}T^\beta\,U_{x,x+\nu}U_{x+\nu,x+\mu+\nu}U_{x+\mu+\nu,x+\mu}U_{x+\mu,x} \nonumber\\
&+& \! \! T^\alpha\, U_{x,x+\nu}U_{x+\nu,x+\nu-\mu}U_{x+\nu-\mu,x-\mu}U_{x-\mu,x} T^\beta\, U_{x,x-\mu}U_{x-\mu,x-\mu+\nu}U_{x-\mu+\nu,x+\nu}U_{x+\nu,x} \nonumber\\
& +& \! \! T^\alpha\, U_{x,x-\mu}U_{x-\mu,x-\mu-\nu}U_{x-\mu-\nu,x-\nu}U_{x-\nu,x} T^\beta\, U_{x,x-\nu}U_{x-\nu,x-\mu-\nu}U_{x-\mu-\nu,x-\mu}U_{x-\mu,x}\nonumber\\
& +& \! \! T^\alpha\, U_{x,x-\nu}U_{x-\nu,x-\nu+\mu}U_{x-\nu+\mu,x+\mu}U_{x+\mu,x} T^\beta\, U_{x,x+\mu}U_{x+\mu,x+\mu-\nu}U_{x+\mu-\nu,x-\nu}U_{x-\nu,x}\bigg).
\eea

The ``Lagrangian mass'', $m_0$, is in principle a free parameter and is related to the bare gluino mass. This term breaks supersymmetry softly. All renormalization factors which we will be calculating, must be evaluated at vanishing renormalized mass, that is, when $m_0$ is set equal to the critical value which ensures a massless gluino in the continuum limit. However, since our calculations are at one loop order this critical value is irrelevant, being already of order $g^2$. The lattice analog of the BRST transformations of the continuum action is shown in Refs.~\cite{Curci:1986sm}, \cite{Kawai:1980ja}.  Further details of the lattice action can be found in Ref.~\cite{Costa:2017rht}. 

Eq.~(\ref{susylagrLattice}) is invariant under the local gauge transformations:
\begin{align}
U'_{x,x+\mu} &= G^{-1}(x) U_{x,x+\mu} G(x+ \mu), &\lambda'(x) &= G^{-1}(x) \lambda(x) G(x), &
\label{SgaugeTranComponentsL}
\end{align}
where $G(x)$ is an element of the $SU(N_c)$ gauge group in the fundamental representation. These gauge transformations commute with the lattice supersymmetry transformations (cf.~\cite{Curci:1986sm}):
\bea
\delta_\xi U_{x,x+\mu} & = & g  \, \bar \xi \gamma_\mu \lambda(x) U_{x,x+\mu}, \nonumber \\
\delta_\xi \lambda(x) & = & \frac{1}{4} [\gamma_{\mu},\gamma_{\nu}] U_{\mu \nu}(x) \xi /(i g) .
\label{susytransfDiracL}
\eea

The lattice discretization of $S_\mu$ is not unique, and alternative variants have been considered in the literature, e.g. ~\cite{Taniguchi:1999fc, Farchioni:2001wx, Bergner:2022wnb}. A definition which we adopt in this work involves stout smeared links and is:
\be
S_\mu = -\frac{1}{2}{\rm tr}_c ( \Tilde{\hat{F}}_{\rho \sigma} [\gamma_\rho,\gamma_\sigma] \gamma_\mu \lambda ),
\quad T_\mu = 2 {\rm tr}_c(\Tilde{\hat{F}}_{\mu \nu} \gamma_\nu \lambda),
\ee
where:
\begin{equation}
\begin{split}
\Tilde{\hat{F}}_{\mu \nu}&=\frac{1}{8ig}(\Tilde{Q}_{\mu \nu}-\Tilde{Q}_{\nu \mu}),\\
\Tilde{Q}_{\mu \nu}&=\Tilde{U}_{x,x+\mu}\Tilde{U}_{x+\mu,x+\mu+\nu}\Tilde{U}_{x+\mu+\nu,x+\nu}\Tilde{U}_{x+\nu,x}\\
\quad& +\Tilde{U}_{x,x+\nu}\Tilde{U}_{x+\nu,x+\nu-\mu}\Tilde{U}_{x+\nu-\mu,x-\mu}\Tilde{U}_{x-\mu,x}\\
\quad& +\Tilde{U}_{x,x-\mu}\Tilde{U}_{x-\mu,x-\mu-\nu}\Tilde{U}_{x-\mu-\nu,x-\nu}\Tilde{U}_{x-\nu,x}\\
\quad& +\Tilde{U}_{x,x-\nu}\Tilde{U}_{x-\nu,x-\nu+\mu}\Tilde{U}_{x-\nu+\mu,x+\mu}\Tilde{U}_{x+\mu,x}.
\end{split}
\end{equation} 

In the present study, we distinguish between the stout smearing parameters used in the fermion action and those employed in the definition of the supercurrent. Specifically, the smearing parameter associated with the operator $S_\mu$, denoted $\omega_O$, is allowed to take a different value from that of the action, denoted $\omega_A$. To maintain generality, both $\omega_A$ and $\omega_O$ are left unspecified throughout the paper, and all expressions are presented in terms of these parameters without assigning them fixed numerical values. The results are presented in analytic form and full generality, with the number of colors $N_c$, the gauge coupling $g$, the gauge-fixing parameter $\alpha$, the clover parameter $c_{\rm SW}$ and the Wilson parameter $r=\pm1$ left unspecified.

Both the $\MSbar$-renormalized and bare Green's functions share the same tensorial structures. As required by renormalizability, the difference between the one-loop $\MSbar$-renormalized Green's functions and the corresponding bare lattice Green's functions [Eqs.~(\ref{latticeq20}), ~(\ref{latticeq10}), (\ref{latticeq2eqmq1}), ~(\ref{3ptGGgLattS})] must be a polynomial in the external momenta. In these differences, one includes only terms that remain nonvanishing in the continuum limit $a \to 0$. Nevertheless, higher-order terms in $a$ are also of significant interest, as they can help reduce lattice artifacts in the corresponding nonperturbative Green's functions. This improvement procedure has been applied in various contexts by our group and others (see, e.g., Refs.~\cite{Capitani:2000xi, Constantinou:2009tr, Gockeler:2010yr, Constantinou:2013ada}). In principle, systematically removing such artifacts could eliminate the need for a continuum extrapolation. However, this is not practically feasible because, without an extrapolation, contributions of the form ${\cal O}(g^m a^n)$ with $m > 2$ and $n > 0$ would eventually become dominant. Terms with $m>4$ are beyond current computational reach. The evaluation of higher-order contributions\footnote{These contributions cannot be extracted analytically for arbitrarily high $n$ as functions of the external momentum $q$. They must instead be computed numerically for each relevant value of $q$.} (${\cal O}(g^2 a^n)$) is a subject for future work and lies outside the scope of the present study.

\smallskip
\subsection{Results for Green's functions and for the mixing matrix in the lattice regularization (LR)}
\label{Latt}

Recently, in Ref.~\cite{Costa:2026zwa}, we computed the renormalizations of the gluino, gluon, and ghost fields, as well as of the coupling constant, in SYM using stout-smeared links; these results will be further employed in the present study. The extraction of the one-loop elements of the mixing matrix involves the difference between the one-loop $\MSbar$-renormalized Green's functions and the corresponding bare lattice Green's functions.

In the evaluation of the Feynman diagrams entering the Green's functions (e.g., Fig.~\ref{fig2pt} for the bare two-point Green's function $\langle u_\nu(q_1)\, S_\mu\, \bar{\lambda}(q_2) \rangle^{B,LR}$), the divergent parts of the loop integrals are independent of the specific lattice discretization (see Ref.~\cite{Costa:2026zwa}), i.e., of the Symanzik coefficients in the gluon action and of the clover and stout-smearing parameters $c_{\rm SW}$, $\omega_A$, and $\omega_O$. These divergent integrals are known in the literature, and we make use of them as in Refs.~\cite{Bergner:2022wnb, Costa:2026zwa}. The remaining convergent parts of the lattice integrals over the loop four-momentum $p_\mu$\,, with $-\pi \le p_\mu \le \pi$, are evaluated by first Taylor-expanding the integrand to the required order in the external momenta. Then we apply a na\"ive discretization of the Brillouin zone, in a way corresponding to a finite hypercubic lattice of size $L$\,: i.e., for each of the four dimensionless integration variables $p_\mu$ of the momentum four-vector, integration over the interval $[0,\,2\pi)$ is converted into a summation over the values $p_{\mu} = 2\pi\,j/L\ (j=0,\ldots,L-1).$ Namely integrals are approximated by a Riemann sum on a hypercubic mesh of $L^4$ points, using several values of $L$ ($4 \le L \le 128$). The results are then extrapolated to the $L\to\infty$ limit. Points corresponding to integrable poles are omitted, since they give a vanishing contribution in this limit. The systematic uncertainty associated with the $L \to \infty$ extrapolation is smaller than the last quoted digit for all results reported in this Subsection. 

Overall, the perturbative calculation on the lattice generates millions of algebraic terms and requires memory and computational resources beyond those of a typical workstation; accordingly, we perform these calculations on high-performance computing (HPC) systems. 

By comparing the lattice Green's function with external gluino, ghost, and antighost fields to the corresponding $\MSbar$-renormalized Green's function in Eq.~\eqref{3ptGFgCCexpS} at one-loop order, one can determine the mixing coefficient $Z_{SA1}^{LR,\MSbar}$:
\be \langle c^R \,{S}^R_{\mu}\, \bar c^R \,\bar\lambda^R \rangle _{amp}\Big|^\MSbar = Z_\lambda^{-1/2} Z_c^{-1} Z_{SS}\langle c^B \,{S}^B_{\mu}\, \bar c^B \,\bar\lambda^B \rangle _{amp} + Z_{SA1} \langle c^B \, {\cal O}^B_{A1} \, \bar c^B\, \bar\lambda^B \rangle _{amp}^{tree} \,  + {\cal O}(g^4). \label{3ptGFexprS} 
\ee 
We find that the two Green's functions coincide at one-loop order in the limit $a\to0$:
\bea 
&&\langle c^{a_3}(q_3) \,S_\mu\,\bar c^{a_2}(q_2) \bar\lambda^{a_1}(q_1) \rangle _{amp}\big|^{\MSbar}_{q_1=q_2,\, q_3 = 0} = \langle c^{a_3}(q_3) \,S_\mu\,\bar c^{a_2}(q_2) \bar\lambda^{a_1}(q_1) \rangle _{amp}\big|^{LR}_{q_1=q_2,\, q_3 = 0}.
\label{GFsCOIN}
\eea 
Since the tree-level Green's function of $S_\mu$ with external gluino, ghost, and antighost fields vanishes, and in view of Eq.~\eqref{GFsCOIN}, we conclude that the mixing of $S_\mu$ with $\mathcal{O}_{A1}$ vanishes:
\be 
Z_{SA1}^{LR,\MSbar} = 0.
\ee

We now turn to the renormalization condition for the two-point Green's function with one external gluon field and one external gluino field, which reads to one loop:
\bea 
\langle u_\nu^R \,{S}^R_{\mu}\, \bar\lambda^R \rangle _{amp}\Big|^\MSbar &=& Z_\la^{-1/2} \,Z_u^{-1/2} Z_{SS} \langle u_\nu^B \,{S}^B_{\mu}\,\bar\lambda^B \rangle _{amp} + Z_{ST}  \langle u_\nu^B \, {T_\mu}^B\, \bar\lambda^B \rangle _{amp}^{tree} + Z_{SA1} \langle  u_\nu^B \, {\cal O}^B_{A1} \, \bar\lambda^B \rangle _{amp}^{tree}\nonumber\\ 
&+& \sum_{i=1}^{2} Z_{SBi} \langle  u_\nu^B \, {\cal O}^B_{Bi} \, \bar\lambda^B \rangle _{amp}^{tree} + \sum_{i=1}^{6} Z_{SCi} \langle u_\nu^B \,{\cal O}^B_{Ci} \, \bar\lambda^B \rangle _{amp}^{tree} + {\cal O}(g^4).
\label{2ptGFexprS}
\eea

Some tree-level two-point Green's functions in Eq.~(\ref{2ptGFexprS}) vanish when one of the external momenta is set to zero, as shown in Table~\ref{tb:treelevels}. This property allows us to disentangle their mixing and determine the mixing coefficients $Z_{SS}, Z_{ST}, Z_{SB_i}$ (with $i=1,2$), $Z_{SC_i}$ (with $i=1,\ldots,6$). The remaining mixing coefficients, $Z_{SC_i}$ ($i=7, 8$), will be extracted from the three-point Green's function with two external gluons and one external gluino.

We first consider the case of vanishing external gluino momentum for $S_\mu$, $q_2 = 0$\,:
\bea
\langle u_\nu^{\alpha_1}(-q_1)  \,{S}_{\mu}\, \bar\lambda^{\alpha_2}(q_2) \rangle _{amp}\big|^{LR}_{q_2 = 0} &=& -i (\slashed{q_1} \gamma_\nu - q_{1\nu}) \gamma_\mu \,\delta^{\alpha_1\,\alpha_2}e^{iq_1x} + i\frac{g^2}{16\pi^2}\frac{1}{2} \delta^{\alpha_1\,\alpha_2}e^{iq_1x}\times \Bigg[ \nonumber \\
&& \hspace{-5cm}
(\gamma_\nu \gamma_\mu \slashed{q_1} + \gamma_\mu q_{1\nu} -2 \gamma_\nu q_{1\mu} ) \bigg(\frac{1}{N_c} (c^{u \bar \la}_{0000} + c^{u \bar \la}_{0001} \omega_O + c^{u \bar \la}_{0002} \omega_O^2) + N_c \Big(c^{u \bar \la}_{1000} + c^{u \bar \la}_{1001} \omega_O + c^{u \bar \la}_{1002} \omega_O^2  \nonumber \\
&& 
+ c^{u \bar \la}_{1010} \omega_A + c^{u \bar \la}_{1011} \omega_A \omega_O + c^{u \bar \la}_{1100} r c_{\rm SW} + c^{u \bar \la}_{1101} r c_{\rm SW} \omega_O + c^{u \bar \la}_{1110} r c_{\rm SW} \omega_A + \nonumber \\ 
&&
 + c^{u \bar \la}_{1111}\,r\, c_{\rm SW} \,\omega_A\, \omega_O +  c^{u \bar \la}_{1200}c_{\rm SW}^2 + c^{u \bar \la}_{1201} c_{\rm SW}^2 \omega_O - 5.1783 \alpha + \frac{1}{2} \alpha^2 \nonumber \\
&&
 - \frac{3}{2}(1-\alpha)\log(a^2 q_1^2) \Big)\bigg) + N_c \bigg( 2\delta_{\nu \mu} \slashed{q_1} - 4 \frac{\slashed{q_1} q_{1\nu} q_{1 \mu}}{q_1^2} + 2 \gamma_\nu q_{1\mu}\bigg) \Bigg],
 \label{latticeq20}
\eea
where the coefficients $c^{u\bar{\lambda}}$ are given in Table~\ref{tb:coeff_culambdaq2zero}. 
\begin{table}[h!]
\centering
\renewcommand{\arraystretch}{1.15}
\begin{tabular}{|c|c|c|c|}
\hline
Gluon action & Wilson & TLS & Iwasaki \\
\hhline{|=|=|=|=|} 
${\rm c}^{u \bar\la}_{0000}$ & -39.478 & -28.919 & -16.602 \\\hline 
${\rm c}^{u \bar\la}_{0001}$ & 631.66 & 449.85 & 248.15 \\ \hline
${\rm c}^{u \bar\la}_{0002}$ & -2842.5 & -1983.6 & -1066.0 \\ \hline
${\rm c}^{u \bar\la}_{1000}$ & 42.074 & 32.347 & 20.425 \\ \hline
${\rm c}^{u \bar\la}_{1001}$ & -586.96 & -422.72 & -237.36 \\ \hline
${\rm c}^{u \bar\la}_{1002}$ & 2368.7 & 1652.4 & 886.53 \\ \hline
${\rm c}^{u \bar\la}_{1010}$ & -4.2355 & -3.5089 & -2.4285 \\ \hline
${\rm c}^{u \bar\la}_{1011}$ & 23.617 & 18.430 & 11.593\\ \hline
${\rm c}^{u \bar\la}_{1100}$ & -5.3771 & -4.8415 & -3.8799 \\ \hline
${\rm c}^{u \bar\la}_{1101}$ & 12.613 & 10.643 & 7.5046 \\ \hline
${\rm c}^{u \bar\la}_{1110}$ & 9.2005 & 7.6316 & 5.2280 \\ \hline
${\rm c}^{u \bar\la}_{1111}$ & -42.090 & -33.603 & -21.490\\ \hline
${\rm c}^{u \bar\la}_{1200}$ & 4.5552 & 4.1497& 3.3963 \\ \hline
${\rm c}^{u \bar\la}_{1201}$ & -9.4356 & -8.0661 & -5.8210 \\ 
\hline
\end{tabular}
\caption{Numerical values of the coefficients ${\rm c}^{u \bar\la}_{ijkl}$ which appear in the contributions ${\rm c}^{u \bar\la}_{ijkl} N_c^{2i-1}\,c_{\rm SW}^j \,\omega_A^k\, \omega_O^l $ for the three actions.}
\label{tb:coeff_culambdaq2zero}
\end{table}

The numerical errors in our one-loop lattice expressions are smaller than the last digit presented. The loop integrals were evaluated numerically using custom routines for lattice momentum sums, implemented in our \href{https://rinnoco.com/projects/super-quanta.html}{SUPER\_QUANTA} software~\cite{Rinnoco:SuperQUANTA, proceedingsLATTICE2026}. Reliability was ensured by standard checks: reproduction of known results, agreement between dimensional and lattice regularizations, gauge-parameter independence where applicable, and consistency with known renormalization structures, including divergence cancellation and correct logarithmic behavior.

The two-point tree-level Green's functions corresponding to the operators $\mathcal{O}_{B1}$, $\mathcal{O}_{B2}$, $\mathcal{O}_{C3}$, and $\mathcal{O}_{C6}$, with external gluon and gluino fields at $q_2 = 0$, vanish. As a result, their mixing coefficients cannot be extracted from this Green's function. However, we are able to determine the renormalization factor $Z_{SS}$, as well as the mixing coefficients for the operators $T_\mu$, $\mathcal{O}_{C1}$, $\mathcal{O}_{C2}$, $\mathcal{O}_{C4}$, and $\mathcal{O}_{C5}$, as shown in Eqs.~(\ref{Zss})--(\ref{ZSA1Cs}). Note that $\mathcal{O}_{C1}$ and $\mathcal{O}_{C2}$ are lower-dimensional operators; they do not appear with inverse powers of the lattice spacing and therefore do not mix at the quantum level. In other words, the absence of $q$-independent terms signals that these lower-dimensional operators do not mix with $S_\mu$. 

In the same vein, for vanishing gluon momentum, $q_1 = 0$, where the tree-level Green's functions for the operators $S_\mu$, $T_\mu$, $\mathcal{O}_{C4}$, and $\mathcal{O}_{C5}$ vanish, we find:
\bea
\langle u_\nu^{\alpha_1}(-q_1)  \,{S}_{\mu}\, \bar\lambda^{\alpha_2}(q_2) \rangle _{amp}\big|^{LR}_{q_1 = 0} &=& i\frac{g^2 N_c}{16\pi^2}\frac{1}{2} \delta^{\alpha_1\,\alpha_2}e^{iq_2x}\times \Bigg[ \left(\frac{1}{2} \gamma_\nu \gamma_\mu \slashed{q_2} + \delta_{\nu \mu} \slashed{q_2} \right) \left(c^{u \bar \la}_{0} + c^{u \bar \la}_{1} \omega_O + \log(a^2 q_2^2) \right) \nonumber \\
&& 
 +\gamma_\nu \gamma_\mu \slashed{q_2}+ \gamma_\mu q_{2 \, \nu} - 2 \gamma_\nu q_{2 \, \mu}  \Bigg],
 \label{latticeq10}
\eea
where the coefficients $c^{u\bar{\lambda}}$ are listed in Table~\ref{tb:coeff_culambdaq1zero}.
\begin{table}[h!]
\centering
\renewcommand{\arraystretch}{1.15}
\begin{tabular}{|c|c|c|c|}
\hline
Gluon action & Wilson & TLS & Iwasaki \\
\hhline{|=|=|=|=|}
${\rm c}^{u \bar\la}_{0}$  & -1.6161 & -1.4111 & -1.0013 \\ \hline
${\rm c}^{u \bar\la}_{1}$  & 4.5777 & 3.9827 & 3.0051 \\ 
\hline
\end{tabular}
\caption{Numerical values of the coefficients ${\rm c}^{u \bar\la}_{0}$, ${\rm c}^{u \bar\la}_{1}$,  for the three actions.}
\label{tb:coeff_culambdaq1zero}
\end{table}

From these Green's functions, one can determine the mixing with the operators that vanish by the equations of motion, $\mathcal{O}_{B1}$ and $\mathcal{O}_{B2}$, as well as $\mathcal{O}_{C3}$ and $\mathcal{O}_{C6}$. The latter two do not contribute to the mixing matrix at one-loop order, even in the presence of stout links.

The last two-point Green's function for $q_2 = -q_1$ is presented below.
\bea
\label{latticeq2eqmq1}
\langle u_\nu^{\alpha_1}(-q_1)  \,{S}_{\mu}\, \bar\lambda^{\alpha_2}(q_2) \rangle _{amp}\big|^{LR}_{q_2 = -q_1} &=& -i (\slashed{q_1} \gamma_\nu - q_{1\nu}) \gamma_\mu \, \delta^{\alpha_1\,\alpha_2}+ i\frac{g^2}{16\pi^2}\frac{1}{2} \delta^{\alpha_1\,\alpha_2}\times \Bigg[ \nonumber \\
&& \hspace{-5cm}
(\gamma_\nu \gamma_\mu \slashed{q_1} + \gamma_\mu q_{1  \nu} -2 \gamma_\nu q_{1 \mu} ) \bigg(\frac{1}{N_c} (c^{u \bar \la}_{0000} + c^{u \bar \la}_{0001} \omega_O + c^{u \bar \la}_{0002} \omega_O^2) + N_c \Big(c^{u \bar \la}_{1000} + c^{u \bar \la}_{1001} \omega_O + c^{u \bar \la}_{1002} \omega_O^2  \nonumber \\
&& 
+ c^{u \bar \la}_{1010} \omega_A + c^{u \bar \la}_{1011} \omega_A \omega_O + c^{u \bar \la}_{1100} r c_{\rm SW} + c^{u \bar \la}_{1101} r c_{\rm SW} \omega_O + c^{u \bar \la}_{1110} r c_{\rm SW} \omega_A + \nonumber \\ 
&&
 + c^{u \bar \la}_{1111}r c_{\rm SW} \omega_A \omega_O +  c^{u \bar \la}_{1200}c_{\rm SW}^2 + c^{u \bar \la}_{1201} c_{\rm SW}^2 \omega_O -6- 4.1783 \alpha + \frac{1}{2} \alpha^2 \nonumber \\
&&
 - \frac{3}{2}(1-\alpha)\log(a^2 q_1^2) \Big)\bigg) - N_c \bigg( (c^{u \bar\la}_{0} + c^{u \bar\la}_{1}\omega_O + \log(a^2 q_1^2))\left(\frac{1}{2}\gamma_\nu \gamma_\mu \slashed{q_{1}} + \delta_{\nu \mu} \slashed{q_1} \right) \nonumber \\
 &&
 +2\alpha(\delta_{\nu \mu} \slashed{q_1} +
  \gamma_\nu q_{1 \mu}) - 4 \frac{\slashed{q_1} q_{1 \nu} q_{1  \mu}}{q_1^2}
 \bigg) \Bigg].
\eea

Eq.~(\ref{latticeq2eqmq1}) provides a nontrivial consistency check, confirming the validity of the results obtained from the previous two-point Green's functions. The coefficients appearing in Eq.~(\ref{latticeq2eqmq1}) are listed in Tables~\ref{tb:coeff_culambdaq2zero} and~\ref{tb:coeff_culambdaq1zero}.

The quantities $Z_{u}^{LR,\overline{\textrm{MS}}}$, $Z_{\lambda}^{LR,\overline{\textrm{MS}}}$, $Z_{g}^{LR,\overline{\textrm{MS}}}$, and $Z_c^{LR,\overline{\textrm{MS}}}$, which appear in the renormalization conditions on the lattice, have been computed in previous works~\cite{Costa:2017rht, Costa:2020keq} at one-loop level for the case $\omega_A = 0$. These quantities have been recalculated in Ref.~\cite{Costa:2026zwa} for the action presented in Eq.~(\ref{susylagrLattice}). It is important to note that, in this case, although the action includes stout-smeared links, the renormalization factors for the gluon, $Z_{u}^{LR,\overline{\textrm{MS}}}$, and ghost fields, $Z_c^{LR,\overline{\textrm{MS}}}$, as well as for the coupling constant, $Z_{g}^{LR,\overline{\textrm{MS}}}$, remain independent of the stout parameter $\omega_A$. Furthermore, since the ghost propagator does not couple to gluino fields at one-loop level, the clover term does not contribute to the one-loop renormalization factor $Z_c^{LR,\overline{\textrm{MS}}}$. In order to provide a self-contained presentation, we list below the renormalization factors of the gluino, gluon, ghost fields and the coupling constant, for the three gluon actions:
\begin{align}
Z_{\lambda}^{LR,\MSbar}
&= 1 - \frac{g^2 N_c}{16\pi^2}\Bigg[
  {\rm{c}}^\la_{00} - 3.7920\,\alpha - {\rm{c}}^\la_{10}\,c_{\rm SW}\,r - {\rm{c}}^\la_{20}\,c_{\rm SW}^2 
  + \alpha \log\!\left(a^2 \bar\mu^2\right)
  - \omega_A\Big( {\rm{c}}^\la_{01} - {\rm{c}}^\la_{11}\,c_{\rm SW}\,r - {\rm{c}}^\la_{02}\,\omega_A \Big)
\Bigg],
\label{ZlaL}
\end{align}
where the coefficients ${\rm{c}}^\lambda_{ij}$ are listed in Table~\ref{tb:coeff_clambda}.
\begin{table}[h!]
\centering
\renewcommand{\arraystretch}{1.15}
\begin{tabular}{|c|c|c|c|}
\hline
Gluon action & Wilson & TLS & Iwasaki \\
\hhline{|=|=|=|=|}
${\rm c}^\la_{00}$ & 16.644 & 13.023 & 8.1166 \\ \hline
${\rm c}^\la_{10}$ & 4.4977 & 4.0309 & 3.2020 \\ \hline
${\rm c}^\la_{20}$ & 5.5891 & 4.9688 & 3.8928 \\ \hline
${\rm c}^\la_{01}$ & 202.01 & 152.56 & 91.144 \\ \hline
${\rm c}^\la_{11}$ & 11.124 & 9.3469 & 6.5475 \\ \hline
${\rm c}^\la_{02}$ & 739.68 & 541.38 & 308.75 \\ 
\hline
\end{tabular}
\caption{Numerical values of the coefficients ${\rm c}^\la_{00},\,{\rm c}^\la_{10},\,{\rm c}^\la_{20},\,{\rm c}^\la_{01},\,{\rm c}^\la_{11},\,{\rm c}^\la_{02}$ for the three actions. Note that at one loop $\alpha$ enters only through the longitudinal gluon propagator, while stout smearing affects only the transverse part, so no $\alpha\times\omega$ term appears.}
\label{tb:coeff_clambda}
\end{table}

\be
Z_{u}^{LR,\MSbar}
= 1 + \frac{g^2}{16\pi^2}
\left[
-\frac{{\rm{c}}_{0}^{u}}{N_c}
- N_c\!\left(
{\rm{c}}_1^{u} - \frac{19}{12}
- 1.3863\,\alpha
- 18.8508\,c_{\rm SW}^2
+ 1.5939\,c_{\rm SW}\,r
+ \left( -\frac{3}{2}+\frac{\alpha}{2}\right) \log\!\left(a^2\bar\mu^2\right)
\right)
\right],
\label{ZuL}
\ee
where the coefficients ${\rm c}_{0}^{u}$ and ${\rm c}_{1}^{u}$ are given in Table~\ref{tb:coeff_u}.
\begin{table}[h!]
\centering
\renewcommand{\arraystretch}{1.15}
\begin{tabular}{|c|c|c|c|}
\hline
Gluon action & Wilson & TLS & Iwasaki \\
\hhline{|=|=|=|=|}
${\rm{c}}^u_{0}$ & -19.739 & -6.6595 & 11.889 \\ \hline
${\rm{c}}^u_{1}$  & 20.147  & 8.1403 & -9.4149 \\
\hline
\end{tabular}
\caption{Numerical values of the coefficients ${\rm{c}}^u_{0}$ and ${\rm{c}}^u_{1}$ for the three gluon actions.}
\label{tb:coeff_u}
\end{table}

\be
Z_c^{LR,\MSbar} = 1 - \frac{g^2 N_c}{16\pi^2} \Bigl[{\rm{c}}^c_{1} - 1 - 1.2029 \, \alpha -\frac{1}{4}\left( 3 - \alpha \right)
\log\left(a^2\,\bar{\mu}^2\right) \Bigr],
\label{ZcL}
\ee
where the coefficient ${\rm c}^{c}_{1}$ is tabulated in Table~\ref{tb:coeff_c}.
\begin{table}[h!]
\centering
\renewcommand{\arraystretch}{1.15}
\begin{tabular}{|c|c|c|c|}
\hline
Gluon action & Wilson & TLS & Iwasaki \\
\hhline{|=|=|=|=|}
${\rm{c}}^c_{1}$ & 4.6086 & 3.7759 & 2.5469 \\
\hline
\end{tabular}
\caption{Numerical values of the coefficient ${\rm{c}}^c_{1}$ for the three gluon actions.}
\label{tb:coeff_c}
\end{table}

\be
Z_g^{LR,\MSbar} =  1 + \gtilde\,\Bigg[\frac{{\rm{c}}_{0}^{u}}{2} \frac{1}{N_c} + N_c \left(\frac{1}{2} \left({\rm{c}}_1^{u} - \frac{19}{12} \right) + {\rm{c}}_1^{c} - 1 - 9.4225 \,c_{\rm SW}^2  + 0.7969 \, c_{\rm SW}\, r - \frac{3}{2} \log\left(a^2\,\bar{\mu}^2\right)\right)\Bigg].
\label{ZgL}
\ee
Eq.~\eqref{ZgL} depends on the coefficients entering $Z_c^{LR,\MSbar}$ and $Z_u^{LR,\MSbar}$.

Starting from the two-point Green's function with the choice $q_2=0$, we obtain the following one-loop results:
\bea
\label{Zss}
Z_{SS}^{LR,\MSbar} = 1 &+& \frac{g^2}{16\pi^2} \Bigg[\frac{1}{N_c}(c^{SS}_{0000} + c^{SS}_{0001} \, \omega_O + c^{SS}_{0002} \, \omega_O^2 ) + N_c \Big(c^{SS}_{1000} + c^{SS}_{1001} \, \omega_O
\nonumber \\
&& +  c^{SS}_{1002} \, \omega_O^2 + c^{SS}_{1010} \, \omega_A + c^{SS}_{1011} \, \omega_A \omega_O + c^{SS}_{1020} \, \omega_A^2 + c^{SS}_{1100} \, r c_{\rm SW}
\nonumber \\
&&+ c^{SS}_{1101} \, r c_{\rm SW} \omega_O + c^{SS}_{1110} \, r c_{\rm SW} \omega_A + c^{SS}_{1111} \, r c_{\rm SW} \omega_A \omega_O + c^{SS}_{1200} \, c_{\rm SW}^2
\nonumber \\
&& + c^{SS}_{1201} \, c_{\rm SW} \omega_O
\Big) \Bigg], \\
\label{Zst}
Z_{ST}^{LR,\MSbar} &=& \frac{g^2}{16\pi^2} 3N_c\,,\\
\label{ZSA1Cs}
Z_{SC4}^{LR,\MSbar}  &=&  Z_{SC5}^{LR,\MSbar}  = 0 .
\eea
The coefficients ${\rm c}^{SS}$ appearing in Eq.~\eqref{Zss} are listed in Table~\ref{tb:coeff_SSculambdaq2zero}.

\begin{table}[h!]
\centering
\renewcommand{\arraystretch}{1.15}
\begin{tabular}{|c|c|c|c|}
\hline
Gluon action & Wilson & TLS & Iwasaki \\
\hhline{|=|=|=|=|}
${\rm c}^{SS}_{0000}$ & -9.8696 & -11.130 & -14.245 \\ \hline
${\rm c}^{SS}_{0001}$ & 315.83 & 224.92 & 124.07 \\ \hline
${\rm c}^{SS}_{0002}$ & -1421.2 & -991.80 & -532.99 \\ \hline
${\rm c}^{SS}_{1000}$ & -2.3170 & 0.63331 & 5.9034 \\ \hline
${\rm c}^{SS}_{1001}$ & -293.48 & -211.36 & -118.69 \\ \hline
${\rm c}^{SS}_{1002}$ & 1184.4 & 826.19 & 443.26 \\ \hline
${\rm c}^{SS}_{1010}$ & 98.886 & 74.528 & 44.358 \\ \hline
${\rm c}^{SS}_{1011}$ & 11.809 & 9.2148 & 5.7965 \\ \hline
${\rm c}^{SS}_{1020}$ & -369.84 & -270.69 & -154.38 \\ \hline
${\rm c}^{SS}_{1100}$ & -1.2366 & -1.2023 & -1.1359 \\ \hline
${\rm c}^{SS}_{1101}$ & 6.3064 & 5.3217 & 3.7523 \\ \hline
${\rm c}^{SS}_{1110}$ & -0.96169 & -0.85764 & -0.65977 \\ \hline
${\rm c}^{SS}_{1111}$ & -21.045 & -16.802 & -10.745 \\ \hline
${\rm c}^{SS}_{1200}$ & 14.498 & 13.985 & 13.070 \\ \hline
${\rm c}^{SS}_{1201}$ & -4.7178 & -4.0331 & -2.9105 \\
\hline
\end{tabular}
\caption{Numerical values of the coefficients ${\rm c}^{SS}_{ijkl}$ which appear in the contributions ${\rm c}^{SS}_{ijkl} N_c^{2i-1}\,c_{\rm SW}^j \,\omega_A^k\, \omega_O^l $ for the three actions. }
\label{tb:coeff_SSculambdaq2zero}
\end{table}

An important feature of the supercurrent operator is that its renormalization is finite: this is in line with its classical conservation. The mixing with $T_\mu$ on the lattice, $Z_{ST}$, is in agreement with Ref.~\cite{Taniguchi:1999fc}, where it is mentioned that $Z_{ST}$ is related to the $\gamma$-trace anomaly corresponding to super-conformal symmetry breaking~\cite{Veneziano:1982ah, Kaymakcalan:1983jh}  and is identical to the one loop level $\beta$-function. 

At this stage, one obtains the gauge-invariant mixing of $S_\mu$ with $T_\mu$. 
Given that the gauge-noninvariant operators do not contribute to physical Green's functions, 
Eq.~(\ref{Zz1}) reduces to:
\begin{equation}
S_\mu^R 
\stackrel{!}{=} Z_{SS}^{LR,\overline{\textrm{MS}}} 
\left( 
S_\mu^B 
+ \frac{Z_{ST}^{LR,\overline{\textrm{MS}}}}{Z_{SS}^{LR,\overline{\textrm{MS}}}} 
T_\mu^B 
\right)
\label{Zz}
\end{equation}
Using Ward identities to renormalize the supercurrent operator naturally leads 
to the appearance of the ratio 
$Z_{ST}^{LR,\overline{\textrm{MS}}}/Z_{SS}^{LR,\overline{\textrm{MS}}}$, 
which has been studied in both perturbative and nonperturbative contexts 
(see, e.g., Refs.~\cite{Farchioni:2001yn, Ali:2018fbq, Bergner:2022see}). 
Our results in Eqs.~(\ref{Zss}) and (\ref{Zst}) are generic and depend on 
parameters such as $g, N_c, r=\pm1, c_{\mathrm{sw}}, \omega_A$, and $\omega_O$, 
making them useful for direct comparison with results from nonperturbative studies. 
It is also worth noting the results of Ref.~\cite{Hieda:2017sqq}, which employed 
the gradient flow and its small flow-time expansion to renormalize the supercurrent.

\bigskip
For the choice $q_2=0$ the tree-level two-point Green's functions of ${\cal O}_{B1}, {\cal O}_{B2}, {\cal O}_{C3}, {\cal O}_{C6}$ vanish; therefore, in order to determine the mixing of these operators, we employ the two-point Green's functions evaluated at $q_1=0$, leading to:
\bea
Z_{SB1}^{LR,\MSbar} &=& \frac{g^2}{16\pi^2} N_c \Bigg[{\rm c}^{SB1}_{0} + {\rm c}^{SB1}_{1} \, \omega_O- \log\left(a^2 \bar \mu^2\right)\Bigg],\label{ZSB1}\\
Z_{SB2}^{LR,\MSbar} &=&  \frac{g^2}{16\pi^2} N_c \Bigg[{\rm c}^{SB2}_{0} + {\rm c}^{SB2}_{1} \, \omega_O -\frac{1}{2} \log\left(a^2 \bar \mu^2\right)\Bigg],\label{ZSB2}\\
Z_{SC3}^{LR,\MSbar} &=& Z_{SC6}^{LR,\MSbar}  = 0.
\eea

The coefficients ${\rm c}^{SB1}_{i}$ and ${\rm c}^{SB2}_{i}$ are listed in Table~\ref{tb:coeff:ZfactorsSB1SB2}.
\begin{table}[h!]
\centering
\renewcommand{\arraystretch}{1.15}
\begin{tabular}{|c|c|c|c|}
\hline
Gluon action & Wilson & TLS & Iwasaki \\
\hhline{|=|=|=|=|}
${\rm{c}}^{SB1}_{0}$ & -0.38395 & -0.58893 & -0.99876 \\ \hline
${\rm{c}}^{SB1}_{1}$ & -4.5777 & -3.9827 & -3.0051 \\ \hline
${\rm{c}}^{SB2}_{0}$ & 0.80802 & 0.70553 & 0.50062 \\ \hline
${\rm{c}}^{SB2}_{1}$ & -2.2889 & -1.9914 & -1.5026 \\
\hline
\end{tabular}
\caption{Numerical values of the coefficients ${\rm{c}}^{SB1}_{0}$, ${\rm{c}}^{SB1}_{1}$, ${\rm{c}}^{SB2}_{0}$, and ${\rm{c}}^{SB2}_{1}$, for the three gluon actions.}
\label{tb:coeff:ZfactorsSB1SB2}
\end{table}

All the above results are consistent with our earlier findings~\cite{Bergner:2022wnb}, 
as verified by comparing the finite parts that do not depend on $\omega_A$ and $\omega_O$, 
as well as the logarithmic divergences in the lattice spacing. At this point, we also 
explicitly check the Green's functions for the choice $q_2 = -q_1$ and find agreement 
with the expressions presented above. 

In order to illustrate graphically the dependence of $Z_{SS}^{LR,\MSbar}$ on the improvement parameters, 
we consider all three gauge actions, setting $N_c=3$ and the Wilson parameter 
to $r=1$. In the figures which follow, we plot the one-loop corrections to $Z_{SS}^{LR,\MSbar} \equiv 1+ \big(g^2/(16\pi^2)\big) \, z_{SS}^{LR,\MSbar}$, omitting the overall coupling-dependent factor $g^2/(16\pi^2)$, as it does not affect the comparison.

\begin{figure}[ht!]

\begin{subfigure}{0.45\textwidth}
\includegraphics[width=0.8\linewidth, height=6cm]{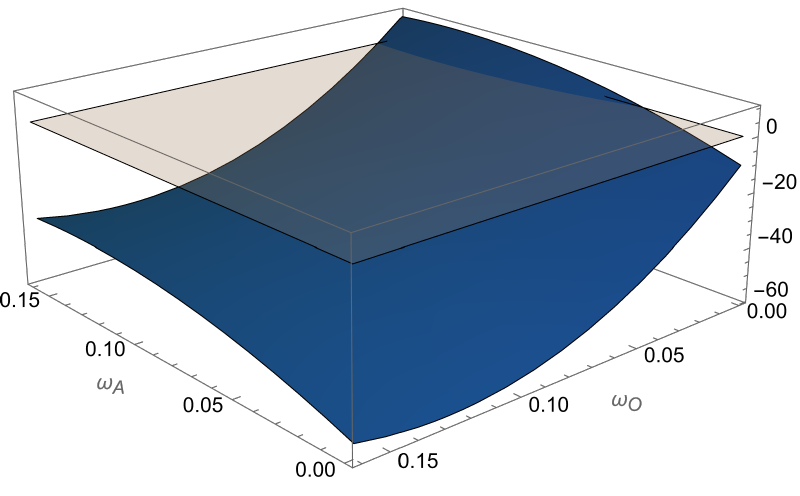} 
\caption{Wilson}
\label{stoutWilson}
\end{subfigure}
\hfill
\begin{subfigure}{0.45\textwidth}
\includegraphics[width=0.8\linewidth, height=6cm]{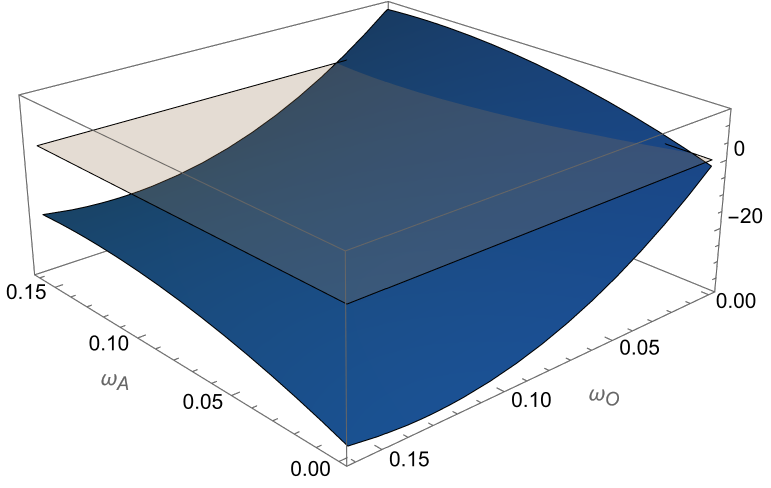}
\caption{TLS}
\label{stoutTLS}
\end{subfigure}
\vspace{1em}
\begin{subfigure}{0.45\textwidth}
\includegraphics[width=0.8\linewidth, height=6cm]{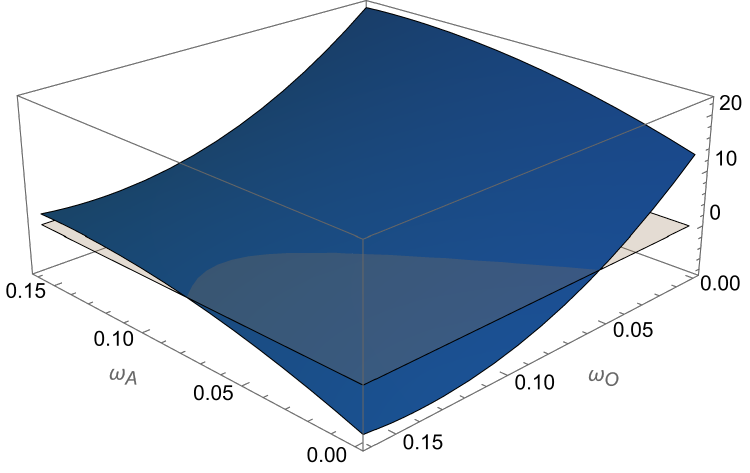}
\caption{Iwasaki}
\label{stoutIwasaki}
\end{subfigure}
\caption{Plots of $z_{SS}^{LR,\MSbar}$ as a function of the stout-smearing parameters $\omega_A$ and $\omega_O$, with the clover coefficient set to $c_{\rm SW}=0$, for each gauge action. The overall factor $g^2/(16\pi^2)$ is omitted. The number of colors is set to $N_c=3$, and the Wilson parameter is fixed at $r=1$. The blue surface represents $z_{SS}^{LR,\MSbar}$, while the semi-transparent plane corresponds to $z_{SS}^{LR,\MSbar}=0$. Both stout-smearing parameters range from $0$ to $0.16$.}
\label{stoutonly}
\end{figure}

We first set $c_{\rm SW}=0$ and allow the stout parameters $\omega_A$ and 
$\omega_O$ to vary independently. The resulting dependence of 
$z_{SS}^{LR,\MSbar}$ on $(\omega_A,\omega_O)$ is shown in 
Fig.~\ref{stoutonly} for the three gauge actions.

\begin{figure}[ht!]

\begin{subfigure}{0.45\textwidth}
\includegraphics[width=0.8\linewidth, height=6cm]{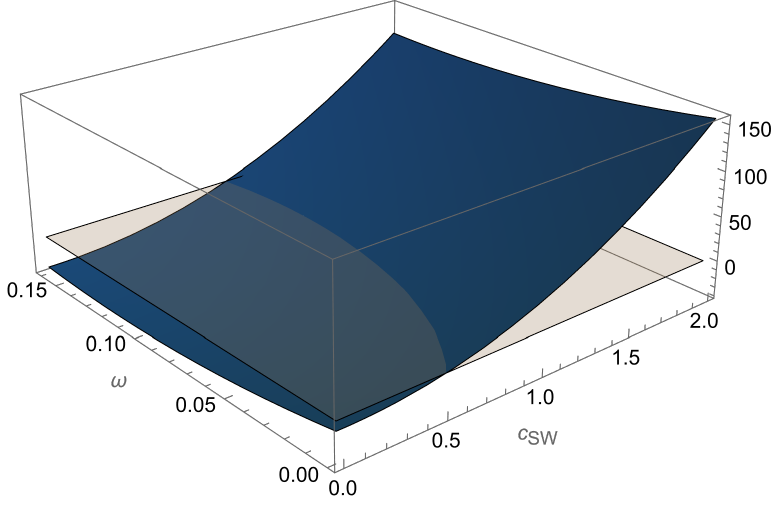} 
\caption{Wilson}
\label{cloverstoutWilson}
\end{subfigure}
\hfill
\begin{subfigure}{0.45\textwidth}
\includegraphics[width=0.8\linewidth, height=6cm]{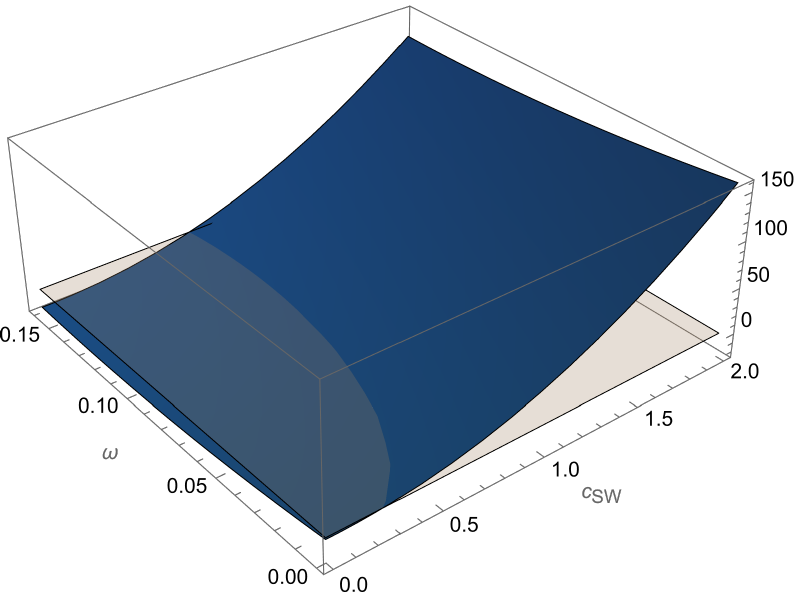}
\caption{TLS}
\label{cloverstoutTLS}
\end{subfigure}
\vspace{1em}
\begin{subfigure}{0.45\textwidth}
\includegraphics[width=0.8\linewidth, height=6cm]{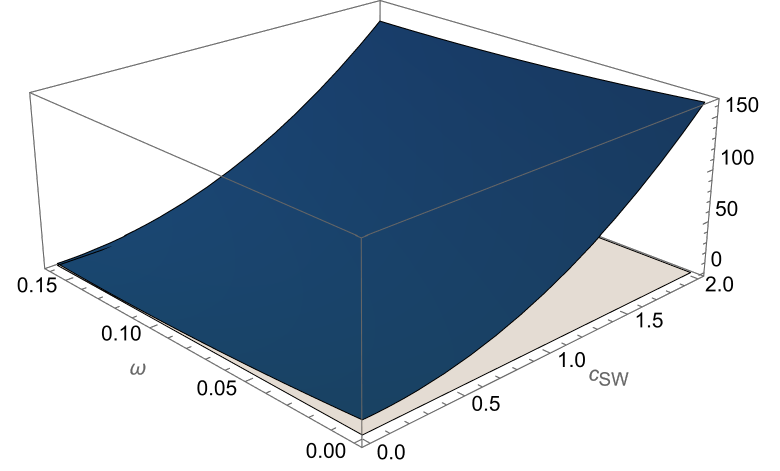}
\caption{Iwasaki}
\label{cloverstoutIwasaki}
\end{subfigure}
\caption{Plots of $z_{SS}^{LR,\MSbar}$ for each gauge action as a function of the clover coefficient $c_{\rm SW}$ and the stout-smearing parameter $\omega$, with the stout-smearing parameters in the fermion action and in the definition of the supercurrent operator set equal ($\omega_A=\omega_O=\omega$). The overall factor $g^2/(16\pi^2)$ is omitted. The number of colors is set to $N_c=3$, and the Wilson parameter is fixed at $r=1$. The blue surface represents $z_{SS}^{LR,\MSbar}$, while the semi-transparent plane corresponds to $z_{SS}^{LR,\MSbar}=0$. The stout-smearing parameter $\omega$ ranges from $0$ to $0.16$, while the clover coefficient $c_{\rm SW}$ ranges from $0$ to $2$.}
\label{cloverstout}
\end{figure}

We then set the two stout parameters equal, 
$\omega_A=\omega_O\equiv\omega$, and allow $c_{\rm SW}$ to vary. 
Figure~\ref{cloverstout} shows the corresponding dependence of 
$z_{SS}^{LR,\MSbar}$ on $(\omega,c_{\rm SW})$ for the three gauge actions.

In Ref.~\cite{Bergner:2022see}, we employed the gauge-invariant renormalization scheme (GIRS) in coordinate space to determine the full $2\times 2$ gauge-invariant mixing matrix $Z$ for the operator basis $\{S_\mu,T_\mu\}$. A total of four independent renormalization conditions are necessary for this purpose. Three conditions were imposed on gauge-invariant two-point functions\footnote{A bar on $S_\mu,T_\mu$ denotes the corresponding charge-conjugate operators.}  of $S_\mu$ and $T_\mu$\,:
\begin{align}
G^{SS}_{\mu\nu}(x,y) &\equiv \langle S_\mu(x)\,\overline{S}_\nu(y)\rangle, \qquad
G^{TT}_{\mu\nu}(x,y) \equiv \langle T_\mu(x)\,\overline{T}_\nu(y)\rangle, \qquad
G^{ST}_{\mu\nu}(x,y) \equiv \langle S_\mu(x)\,\overline{T}_\nu(y)\rangle,
\label{eq:GFs_short}
\end{align}
and we used a fourth condition involving the only other gauge-invariant operator of equal or lower dimension with compatible Lorentz structure, the gluino--glue operator:
\begin{equation}
\mathcal{O}(x)\equiv \sigma_{\rho\sigma}\,\mathrm{tr}_c\!\big(u_{\rho\sigma}(x)\lambda(x)\big),\qquad
G^{\mathcal{O}S}_{\mu}(x,y)\equiv \langle \mathcal{O}(x)\,\overline{S}_\mu(y)\rangle.
\label{eq:O_def_short}
\end{equation}

For improved numerical signal, we integrated over the spatial separation $z=y-x=(\vec z,t)$ at fixed Euclidean time $t$, and imposed, for $AB\in\{SS,TT,ST\}$,
\begin{align}
\int d^3\vec z\, \mathrm{Tr}\!\left\{\left[G^{AB}_{\mu\nu}(x,y)\right]^{\mathrm{GIRS}} P_{\nu\mu}\right\}
&=
\int d^3\vec z\, \mathrm{Tr}\!\left\{\left[G^{AB}_{\mu\nu}(x,y)\right]^{\mathrm{tree}} P_{\nu\mu}\right\},
\label{eq:GIRS_conds_short}
\\
\int d^3\vec z\, \mathrm{Tr}\!\left\{\left[G^{\mathcal{O}S}_{\mu}(x,y)\right]^{\mathrm{GIRS}} P_{\mu}\right\}
&=
\int d^3\vec z\, \mathrm{Tr}\!\left\{\left[G^{\mathcal{O}S}_{\mu}(x,y)\right]^{\mathrm{tree}} P_{\mu}\right\}.
\label{eq:GIRS_cond_O_short}
\end{align}
The Dirac projectors were chosen as $P_{\nu\mu}=\gamma_\mu\gamma_4\gamma_\nu$ and $P_\mu=\gamma_\mu\gamma_4$. Repeated indices were not summed over; the same $(\mu,\nu)$ had to be used in all conditions since, in principle, different components could have led to different renormalization factors in this scheme. As shown in Ref.~\cite{Bergner:2022see}, choosing a temporal index could have yielded vanishing tree-level normalizations and led to ill-defined solutions; hence we restricted to purely spatial components for $\mu,\nu$.

The explicit tree-level normalizations after spatial integration, and the resulting (quadratic) system determining the matrix elements of $Z$ (and $Z_{\mathcal O}$), were given in Ref.~\cite{Bergner:2022see}.

Next, we evaluate the ratio $Z_{ST}^{LR,\MSbar}/Z_{SS}^{LR,\MSbar}$ 
numerically. We consider the gauge groups 
${\rm SU}(2)$ and ${\rm SU}(3)$. The Wilson parameter $r$ is set to $1$, and
$g^2=2N_c/\beta$, with $\beta=1.75$ for ${\rm SU}(2)$ and 
$\beta=5.6$ for ${\rm SU}(3)$. For ${\rm SU}(2)$ we take 
$c_{\rm SW}=0$, while for ${\rm SU}(3)$ we consider the following set of values
$c_{\rm SW}=0,1,1.587$. The values of the different parameters were chosen in accordance to Ref.~\cite{Bergner:2022see}. Results are presented both without stout smearing 
and with smearing applied only to the operator, corresponding to 
$\omega_A=0$ and $\omega_O=0.15$.

The numerical results for the Wilson, tree-level Symanzik (TLS), and 
Iwasaki gauge actions are summarized in Table~\ref{tb:ZSSZST}. For the 
Wilson action without stout smearing, our results are in full agreement 
with Ref.~\cite{Bergner:2022see}.

\begin{table}[h!]
\centering
\renewcommand{\arraystretch}{1.15}
\begin{tabular}{|c|c|c|c|c|c|}
\hline
Gauge group & $c_{\rm SW}$ & Stout & Wilson & TLS & Iwasaki \\
\hhline{|=|=|=|=|=|=|}
${\rm SU}(2)$ & 0 & No  & 0.10081  &  0.092608 & 0.081332 \\ \hline
${\rm SU}(2)$ & 0 & Yes & 0.18479  & 0.13553 & 0.097682 \\ \hline
${\rm SU}(3)$ & 0 & No  & 0.065624 & 0.061823 & 0.056128 \\ \hline
${\rm SU}(3)$ & 0 & Yes & 0.099827 & 0.081788 & 0.064716 \\ \hline
${\rm SU}(3)$ & 1 & No  & 0.050868 & 0.048934 & 0.045883 \\ \hline
${\rm SU}(3)$ & 1 & Yes & 0.068885 & 0.060416 & 0.051355 \\ \hline
${\rm SU}(3)$ & 1.587 & No  & 0.037376 & 0.036658 & 0.035476 \\ \hline
${\rm SU}(3)$ & 1.587 & Yes & 0.046641 & 0.043019 & 0.038828 \\
\hline
\end{tabular}
\caption{Numerical values of the ratio 
$Z_{ST}^{LR,\MSbar}/Z_{SS}^{LR,\MSbar}$ for the Wilson, 
tree-level Symanzik (TLS), and Iwasaki gauge actions. For the 
stout-smeared results, smearing is applied only to the operator, 
with $\omega_A=0$ and $\omega_O=0.15$. The values of $\beta$ used for the gauge groups SU$(2)$ and SU$(3)$ were $\beta=1.75$ and $\beta=5.6$ respectively.}
\label{tb:ZSSZST}
\end{table}

\medskip
The determination of the mixing with ${\cal O}_{C7}$ and ${\cal O}_{C8}$ requires the evaluation of three-point Green's functions. The Green's function with two external gluons and one external gluino is shown below for a specific choice of external momenta:

\bea
&& \langle u_\nu^{\alpha_1}(-q_1) u_\rho^{\alpha_2}(-q_2)
\,{S}_{\mu}\,\bar\lambda^{\alpha_3}(q_3) \rangle _{amp}
\big|^{LR}_{q_2=0,\,q_3=-q_1}
=
g\,\frac{[\gamma_\nu,\gamma_\rho]\gamma_\mu}{2}
+\frac{g^3}{16\pi^2}\,
f^{\alpha_1\alpha_2\alpha_3}
\Bigg\{
\left(
\delta_{\nu\rho}\gamma_\mu
-\gamma_\nu\gamma_\rho\gamma_\mu
\right)\times \nonumber\\
&&\hspace{0.5cm}
\Bigg[
\frac{1}{N_c}
\left(
c_{0000}^{uu\bar\la}
+c_{0001}^{uu\bar\la}\,\omega_O
+c_{0002}^{uu\bar\la}\,\omega_O^2
\right)
+N_c\bigg(
c_{1000}^{uu\bar\la}
+c_{1001}^{uu\bar\la}\,\omega_O
+c_{1002}^{uu\bar\la}\,\omega_O^2
+c_{1010}^{uu\bar\la}\,\omega_A
+c_{1011}^{uu\bar\la}\,\omega_A\omega_O
+c_{1100}^{uu\bar\la}\,rc_{\rm SW}
\nonumber\\
&&\hspace{0.5cm}
+c_{1101}^{uu\bar\la}\,rc_{\rm SW}\omega_O
+c_{1110}^{uu\bar\la}\,rc_{\rm SW}\omega_A
+c_{1111}^{uu\bar\la}\,rc_{\rm SW}\omega_A\omega_O
+c_{1200}^{uu\bar\la}\,c_{\rm SW}^2
+c_{1201}^{uu\bar\la}\,c_{\rm SW}^2\omega_O
-3.0323\,\alpha
+\frac{\alpha^2}{4}
\nonumber\\
&&\hspace{0.5cm}
-\frac{1-2\alpha}{2}\log(a^2q_1^2)
\bigg)
\Bigg]
+N_c\bigg(
\delta_{\nu\rho}\gamma_\mu
\left(\frac{11}{16}+\frac{3\alpha}{8}\right)
-\delta_{\nu\mu}\gamma_\rho
\left(\frac{3}{8}+\frac{3\alpha}{4}\right)
+\delta_{\rho\mu}\gamma_\nu
\left(\frac{3}{4}-\frac{\alpha}{2}\right)
\nonumber\\
&&\hspace{0.5cm}
- \gamma_\nu \gamma_\mu \frac{\slashed{q_1} q_{1\rho}}{q^2}
\left( \frac{21}{16} -\frac{13\alpha}{8} -\frac{\alpha^2}{4}\right)
+ \gamma_\nu \gamma_\rho \frac{\slashed{q_1} q_{1\mu}}{q^2}
\left( \frac{9}{8} +\frac{\alpha}{4} \right)
+ \gamma_\rho \gamma_\mu \frac{\slashed{q_1} q_{1\nu}}{q^2}
\left( \frac{11}{16} +\frac{\alpha}{4} \right) \nonumber\\
&&\hspace{0.5cm}
- \delta_{\nu \mu} \frac{\slashed{q_1} q_{1\rho}}{q_1^2}
\left(\frac{7}{8} +\frac{3\alpha}{4} \right)
+ \delta_{\nu \rho} \frac{\slashed{q_1} q_{1\mu}}{q_1^2}
\left(\frac{5}{4} +\frac{5\alpha}{2} \right)
- \delta_{\mu \rho} \frac{\slashed{q_1} q_{1\nu}}{q_1^2}
\left(\frac{5}{8} - \frac{3\alpha}{2} \right) \nonumber\\
&&\hspace{0.5cm}
+ \gamma_\mu \frac{q_{1\nu} q_{1\rho}}{q_1^2}
\left(\frac{3}{8} +\frac{9\alpha}{4}+ \frac{\alpha^2}{4} \right)
- \gamma_\nu \frac{q_{1\rho} q_{1\mu}}{q_1^2}
\left(\frac{1}{2} +\frac{9\alpha}{2}+\frac{\alpha^2}{2} \right)
+ \gamma_\rho \frac{q_{1\nu} q_{1\mu}}{q_1^2}
\left(\frac{1}{4} +\frac{3\alpha}{2} \right) \nonumber\\
&&\hspace{0.5cm}
- (1 + 3\alpha)
\frac{\slashed{q_1} q_{1\nu} q_{1\rho} q_{1\mu}}{q_{1}^4}
\bigg)
\Bigg\}.
\label{3ptGGgLattS}
\eea

where the coefficients ${\rm c}^{uu\bar{\lambda}}$ are listed in Table~\ref{tb:coeff_cuulambda}.

\begin{table}[h!]
\centering
\renewcommand{\arraystretch}{1.15}
\begin{tabular}{|c|c|c|c|}
\hline
Gluon action & Wilson & TLS & Iwasaki \\
\hhline{|=|=|=|=|} 
${\rm c}^{uu \bar\la}_{0000}$ & -19.739 & -14.460 & -8.3010 \\\hline 
${\rm c}^{uu \bar\la}_{0001}$ & 315.83 & 224.92 & 124.07 \\ \hline
${\rm c}^{u u\bar\la}_{0002}$ & -1421.2 & -991.80 & -532.99 \\ \hline
${\rm c}^{uu \bar\la}_{1000}$ & 14.987 & 10.853 & 5.9163 \\ \hline
${\rm c}^{uu \bar\la}_{1001}$ & -295.77 & -213.35 & -120.18 \\ \hline
${\rm c}^{uu \bar\la}_{1002}$ & 1184.4 & 826.19 & 443.26 \\ \hline
${\rm c}^{uu \bar\la}_{1010}$ & -2.1177 & -1.7545 & -1.2142 \\ \hline
${\rm c}^{uu \bar\la}_{1011}$ & 11.809 & 9.2148 & 5.7965 \\ \hline
${\rm c}^{uu \bar\la}_{1100}$ & -2.6885 & -2.4208 & -1.9400 \\ \hline
${\rm c}^{uu \bar\la}_{1101}$ & 6.3064 & 5.3217 & 3.7523 \\ \hline
${\rm c}^{uu \bar\la}_{1110}$ & 4.6002 & 3.8158 & 2.6140 \\ \hline
${\rm c}^{u u\bar\la}_{1111}$ & -21.045 & -16.802 & -10.745 \\ \hline
${\rm c}^{u u\bar\la}_{1200}$ & 2.2776 & 2.0748 & 1.6981 \\ \hline
${\rm c}^{u u\bar\la}_{1201}$ & -4.7178 & -4.0331 & -2.9105 \\ \hline
\end{tabular}
\caption{Numerical values of the coefficients ${\rm c}^{uu \bar\la}_{ijkl}$ which appear in the contributions ${\rm c}^{u u\bar\la}_{ijkl} N_c^{2i-1}\,c_{\rm SW}^j \,\omega_A^k\, \omega_O^l $ for the three actions.}
\label{tb:coeff_cuulambda}
\end{table}

Following the same procedure for extracting the mixing of $S_\mu$ with ${\cal O}_{C7}$ and ${\cal O}_{C8}$, we use Eq. \ref{3ptGGgLattS}:

\bea
\langle u_\nu^R u_\rho^R \,{S}^R_{\mu}\, \bar\lambda^R \rangle _{amp} \Big|^\MSbar &=&Z_\la^{-1/2} \,Z_u^{-1}\, Z_{SS} \langle u_\nu^B u_\rho^B \,{S}^B_{\mu}\,\bar\lambda^B \rangle _{amp} + Z_{ST} \langle u_\nu^B u_\rho^B \, {T_\mu}^B\, \bar\lambda^B \rangle _{amp}^{tree} \nonumber\\
&+&  Z_{SB1}\langle u_\nu^B u_\rho^B \, {\cal O}^B_{B1}\, \bar\lambda^B \rangle _{amp}^{tree} + Z_{SB2}\langle u_\nu^B u_\rho^B \, {\cal O}^B_{B2}\, \bar\lambda^B \rangle _{amp}^{tree} \nonumber\\
&+& \sum_{i=7}^{8} Z_{SCi}\langle u_\nu^B u_\rho^B \, {\cal O}^B_{Ci}\, \bar\lambda^B \rangle _{amp}^{tree} + {\cal O}(g^4).
\label{3ptGFexpr1S}
\eea

We find that there is no mixing with these operators to one loop:
\bea
Z_{SC7}^{LR,\MSbar} &=& Z_{SC8}^{LR,\MSbar} = 0.
\eea

Thus, at one-loop order on the lattice, $S_\mu$ does not mix with class-A or class-C operators. It does, however, mix with the gauge-variant operators of class B; cf. Eqs.~\eqref{ZSB1}--\eqref{ZSB2}. Somewhat unexpectedly, the introduction of stout-smeared links does not alter the pattern of operator mixing: the same class-B operators contribute as in the unsmeared case. Nevertheless, our results demonstrate that the stout-smearing parameter can be appropriately tuned to suppress the corresponding mixing coefficients, thereby bringing the theory closer to the supersymmetric point, where the multiplicative renormalization factor is unity, i.e., $Z_{SS}^{LR,\MSbar}=1$.

\section{Summary and Future Plans}
\label{summary}

In this work, we perform a one-loop perturbative calculation of two-point and three-point Green's functions involving the Noether supercurrent operator $S_\mu$ in $\mathcal{N}=1$ Supersymmetric Yang--Mills (SYM) theory. The lattice perturbative calculations are carried out using stout-smeared gauge links, which improve the renormalization behavior of the operators by suppressing ultraviolet fluctuations and reducing discretization artifacts, while preserving gauge invariance. In addition, we employ Symanzik-improved gluons and present results for the three commonly used gluon actions. The resulting Green's functions enable the extraction of renormalization constants and mixing coefficients required for a consistent definition of renormalized lattice operators. While stout smearing and Symanzik improvement lead to improved convergence properties, they also substantially increase the complexity of the interaction vertices, making the evaluation of the corresponding Feynman diagrams significantly more demanding.

In Ref.~\cite{Bergner:2022see}, we presented nonperturbative results obtained in the Gauge-Invariant Renormalization Scheme (GIRS), along with conversion factors that relate GIRS to the $\overline{\text{MS}}$ scheme. These results constitute a foundational step toward a nonperturbative evaluation of supersymmetric Ward identities. A precise determination of renormalization and mixing coefficients will be important in future numerical simulations aimed at testing supersymmetric Ward identities, investigating the restoration of supersymmetry in the continuum limit, and exploring the nonperturbative dynamics of $\mathcal{N}=1$ SYM theory.

Future work will focus on a detailed study of the ``gluino-glue'' operator within the same discretization framework, combining stout-smeared gauge links with Symanzik-improved gauge actions. In particular, we plan to determine its renormalization and possible operator mixing both perturbatively and nonperturbatively, and to assess the impact of smearing and improvement on the associated lattice artifacts. These developments will further support high-precision numerical investigations of supersymmetric Ward identities and provide additional insight into the low-energy nonperturbative structure of $\mathcal{N}=1$ SYM theory, in particular its bound states.

\begin{acknowledgments}
 The projects EXCELLENCE/0524/0208 and EXCELLENCE/0524/0001 are implemented under the programme of social cohesion ``THALIA 2021-2027'' co-funded by the European Union, through the Research and Innovation Foundation of Cyprus. We thank Gregoris Spanoudes for helpful discussions on the stout smearing of the supercurrent operator.
\end{acknowledgments}

\newpage
\appendix

\section{Results for $\MSbar$-renormalized Green's functions of $S_\mu$}
\label{app:MSbarGreenFunctions}

In this appendix, we present the $\MSbar$-renormalized Green's functions of the supercurrent operator $S_\mu$ up to one-loop order. These results are derived using dimensional regularization, with the $1/\epsilon$ divergences subtracted from the corresponding bare Green's functions provided in Ref.~\cite{Bergner:2022wnb}. For completeness, all relevant expressions are included here, as they constitute essential components for matching with lattice-regularized quantities and for determining the renormalization and mixing coefficients, $Z^{LR,\MSbar}$, within the lattice formulation.

The two-point $\MSbar$-renormalized Green's function of $S_\mu$ for $q_2 = 0$ with external gluon and gluino fields are:
 \bea
 \langle u_\nu^{\alpha_1}(-q_1)  \,{S}_{\mu}\, \bar\lambda^{\alpha_2}(q_2) \rangle _{amp}\big|^{\MSbar}_{q_2 = 0}  &=& -i\delta^{\alpha_{1}\alpha_{2}} e^{iq_{1}x} (\slashed q_{1}\gamma_{\nu}-q_{1\nu})\gamma_\mu + i\frac{g^2 N_{c}}{16\pi^2}\frac{1}{2} \delta^{\alpha_1\,\alpha_2}e^{iq_1x}\nonumber \\ &\times& 
\bigg[
(\gamma_{\nu} \gamma_{\mu} \slashed q_1 + \gamma_\mu q_{1\nu})\left(\frac{23}{2} + \frac{\alpha^2}{2}  + \frac{3 (1-\alpha)}{2} \log\left(\frac{\bar \mu^2}{q_1^2}\right)\right) \nonumber \\
&&- \gamma_{\nu} q_{1\mu}\left(15 + \alpha^2 + 3(1- \alpha) \log\left(\frac{\bar \mu^2}{q_1^2}\right)\right) \nonumber \\
&& - 4 \left(\frac{\slashed q_{1} q_{1 \mu} q_{1 \nu}}{q_1^2} +\slashed q_{1} \delta_{\mu\,\nu} \right)
\bigg]
\label{GgCONTq2zero}
\eea

In the case of $q_1 = 0$, we find:
\bea
\langle u_\nu^{\alpha_1}(-q_1)  \,{S}_{\mu}\, \bar\lambda^{\alpha_2}(q_2) \rangle _{amp}\big|^{\MSbar}_{q_1 = 0}  &=&  i\frac{g^2 N_{c}}{16\pi^2}\frac{1}{2} \delta^{\alpha_1\,\alpha_2}e^{iq_2x}\nonumber \\ &\times& \bigg[
\gamma_{\nu} \gamma_{\mu} \slashed q_2 \left(1 - \frac{1}{2} \log\left(\frac{\bar \mu^2}{q_2^2}\right)\right) \nonumber \\
&&+ \gamma_\mu q_{2\nu} - 2 \gamma_{\nu} q_{2\mu}  - \slashed q_{2} \delta_{\mu\,\nu}\left(2 + \log\left(\frac{\bar \mu^2}{q_2^2}\right)\right)
\bigg]
\label{GgCONTq1zero}
\eea

The two-point $\MSbar$-renormalized Green's function of $S_\mu$ for $q_2 = -q_1$ with external gluon and gluino fields, takes the following form:
\bea
 \langle u_\nu^{\alpha_1}(-q_1)  \,{S}_{\mu}\, \bar\lambda^{\alpha_2}(q_2) \rangle _{amp}\big|^{\MSbar}_{q_2 = -q_1}  &=& -i\delta^{\alpha_{1}\alpha_{2}} (\slashed q_{1}\gamma_{\nu}-q_{1\nu})\gamma_\mu  + i\frac{g^2 N_{c}}{16\pi^2}\frac{1}{2} \delta^{\alpha_1\,\alpha_2} \nonumber \\ &\times& \bigg[
 \gamma_{\nu} \gamma_{\mu} \slashed q_1 \left(\frac{11}{2} + \alpha + \frac{\alpha^2}{2}  + \frac{4 - 3 \alpha }{2} \log\left(\frac{\bar \mu^2}{q_1^2}\right)\right) \nonumber \\
 && + \gamma_\mu q_{1\nu} \left( -5  - 4 \alpha -\alpha^2 +  3 (-1+\alpha)  \log\left(\frac{\bar \mu^2}{q_1^2}\right)\right) \nonumber \\
&& + \slashed q_{1} \delta_{\mu\,\nu} \left(-4 - 2\alpha +\log\left(\frac{\bar \mu^2}{q_1^2}\right)\right) \nonumber \\
&&- \gamma_{\nu} q_{1\mu}\left(5 + 4 \alpha + \alpha^2 + 3(1- \alpha) \log\left(\frac{\bar \mu^2}{q_1^2}\right)\right) \nonumber \\
&& + 4 \alpha \frac{\slashed q_{1} q_{1 \mu} q_{1 \nu}}{q_1^2}
\bigg]
\label{GgCONTq2Pq1zero}
\eea

Further, we present the three-point $\MSbar$-renormalized Green's functions of $S_\mu$, with two external gluon fields and one gluino field, for the specific choice of the external momenta for the gluon and gluino fields ($q_2 = 0$, $q_3 = - q_1$)
\footnote{Note the presence of $Z_g$, which is required at one loop in this case and must be taken into account in order to write the $\MSbar$-renormalized Green's Functions}:
\begin{eqnarray}
\langle u_\nu^{\alpha_1}(-q_1) u_\rho^{\alpha_2}(-q_2) \,{S}_{\mu}\, \bar\lambda^{\alpha_3}(q_3) \rangle _{amp}\big|^{DR}_{q_2 =0 , q_3 = -q_1} &=& 
 \, g^R \,f^{\alpha_{1}\alpha_{2}\alpha_{3}} \frac{1}{2}[\gamma_\nu,\gamma_\rho] \,\gamma_\mu \, Z_{g}^{-1}\nonumber\\
&&\hspace{-5cm}- \frac{(g^R)^3 N_c}{16\pi^2} \,f^{\alpha_{1}\alpha_{2}\alpha_{3}}\, \Bigg[
\gamma_\nu \gamma_\rho \gamma_\mu \left(\frac{1-2\alpha}{2 \epsilon} +  \frac{5}{2} +\frac{\alpha}{4} +\frac{\alpha^2}{4} +\frac{1-2\alpha}{2} \log \left(\frac{\bar \mu^2}{q_1^2}\right)\right) \nonumber\\
&&\hspace{-5cm}+ \delta_{\nu \rho} \gamma_\mu \left(-\frac{1-2\alpha}{2 \epsilon} - \frac{29}{16} +\frac{\alpha}{8} - \frac{\alpha^2}{4} - \frac{1-2\alpha}{2} \log \left(\frac{\bar \mu^2}{q_1^2}\right)\right) 
+ \delta_{\nu \mu} \gamma_\rho \left(\frac{19}{8} +\frac{3\alpha}{4} \right) 
+ \delta_{\mu \rho} \gamma_\nu \left(- \frac{11}{4} +\frac{\alpha}{2} \right) \nonumber\\
&&\hspace{-5cm}+ \gamma_\nu \gamma_\mu  \frac{\slashed{q_1} q_{1\rho}}{q^2} \left( \frac{21}{16} -\frac{13\alpha}{8} -\frac{\alpha^2}{4}\right)
- \gamma_\nu \gamma_\rho  \frac{\slashed{q_1} q_{1\mu}}{q^2} \left( \frac{9}{8} +\frac{\alpha}{4} \right) 
               - \gamma_\rho \gamma_\mu  \frac{\slashed{q_1} q_{1\nu}}{q^2} \left( \frac{11}{16} +\frac{\alpha}{4} \right) \nonumber\\
&&\hspace{-5cm}+  \delta_{\nu \mu} \frac{\slashed{q_1} q_{1\rho}}{q_1^2} \left(\frac{7}{8} +\frac{3\alpha}{4} \right)
               -  \delta_{\nu \rho} \frac{\slashed{q_1} q_{1\mu}}{q_1^2} \left(\frac{5}{4} +\frac{5\alpha}{2} \right)
               + \delta_{\mu \rho} \frac{\slashed{q_1} q_{1\nu}}{q_1^2} \left(\frac{5}{8} - \frac{3\alpha}{2} \right) \nonumber\\
&&\hspace{-5cm} - \gamma_\mu \frac{q_{1\nu} q_{1\rho}}{q_1^2} \left(\frac{3}{8} +\frac{9\alpha}{4}+ \frac{\alpha^2}{4} \right) 
+ \gamma_\nu \frac{q_{1\rho} q_{1\mu}}{q_1^2} \left(\frac{1}{2} +\frac{9\alpha}{2}+\frac{\alpha^2}{2} \right)
- \gamma_\rho \frac{q_{1\nu} q_{1\mu}}{q_1^2} \left(\frac{1}{4} +\frac{3\alpha}{2} \right) \nonumber\\
+ (1 + 3\alpha)\frac{\slashed{q_1} q_{1\nu} q_{1\rho} q_{1\mu}}{q_{1}^4} 
\Bigg]
\label{threeuulS}
\end{eqnarray}

The three-point Green’s function involving $S_\mu$, with external momenta corresponding to the gluino and the ghost/antighost fields, is finite in dimensional regularization (DR) and therefore coincide with its $\MSbar$ counterpart. For the choices of the external momenta where $q_1 = q_2$ and $q_3 = 0$, it takes the form:
\bea
\langle c^{a_3}(q_3) \,S_\mu\,\bar c^{a_2}(q_2) \bar\lambda^{a_1}(q_1) \rangle _{amp}\big|^{\MSbar}_{q_1=q_2,\, q_3 = 0}= -\frac{g^3 N_{c}}{16\pi^2}  \,f^{\alpha_{1}\alpha_{2}\alpha_{3}}\, \alpha \left(\frac{1}{4} \gamma_\mu - \frac{\slashed{q_1} q_{1\mu}}{q_1^2}\,\right) 
\label{3ptGFgCCexpS}
\eea
Given that this Green's function vanishes at tree level and is finite at one-loop order, this confirms the absence of mixing with the operator $\mathcal{O}_{A1}$ in the $\MSbar$ scheme. However, on the lattice, the corresponding mixing coefficient $Z_{SA1}$ may receive finite contributions.

\end{document}